\documentclass[journal]{IEEEtran}
\usepackage{graphicx}      
\usepackage{subfig}
\usepackage[square,sort,comma,numbers]{natbib}
\usepackage{amsmath}
\usepackage{amsfonts}
\usepackage{tablefootnote}
\usepackage{url}
\usepackage[table]{xcolor}
\usepackage{color}
\usepackage{caption}
\usepackage{bm}
\usepackage{xcolor}

\usepackage{nth}
\usepackage{tikz}
\usepackage{verbatim}
\usepackage{tkz-euclide}

\definecolor{ao(english)}{rgb}{0.0, 0.5, 0.0}

\usepackage[utf8]{inputenc}

\newenvironment{rcases}
  {\left.\begin{aligned}}
  {\end{aligned}\right\rbrace}

\begin{document}
%
\title{A modified sequence domain impedance definition and its equivalence to the dq-domain impedance definition for the stability analysis of AC power electronic systems \\}

\author{{Atle~Rygg,~Marta~Molinas,~Zhang~Chen~and~Xu~Cai}%

\thanks{A. Rygg and M. Molinas are with the Department of Engineering Cybernetics, Norwegian University of Science and Technology, O. S. Bragstads plass 2D, 7032 Trondheim, Norway, e-mail: atle.rygg@itk.ntnu.no, tel: +47 95977295}%
\thanks{Z. Chen and X. Cai are with the State Energy Smart Grid R\&D Center, Shanghai Jiao Tong University, Shanghai, China, e-mail: zhangchencumt@163.com}}

\maketitle

\begin{abstract}             
Representations of AC power systems by frequency dependent impedance equivalents is an emerging technique in the dynamic analysis of power systems including power electronic converters. The technique has been applied for decades in DC-power systems, and it was recently adopted to map the impedances in AC systems. Most of the work on AC systems can be categorized in two approaches. One is the analysis of the system in the \textit{dq}-domain, whereas the other applies harmonic linearization in the phase domain through symmetric components. Impedance models based on analytical calculations, numerical simulation and experimental studies have been previously developed and verified in both domains independently. The authors of previous studies discuss the advantages and disadvantages of each domain separately, but neither a rigorous comparison nor an attempt to bridge them has been conducted. The present paper attempts to close this gap by deriving the mathematical formulation that shows the equivalence between the \textit{dq}-domain and the sequence domain impedances. A modified form of the sequence domain impedance matrix is proposed, and with this definition the stability estimates obtained with the Generalized Nyquist Criterion (GNC) become equivalent in both domains. The second contribution of the paper is the definition of a \textit{Mirror Frequency Decoupled} (MFD) system. 
The analysis of MFD systems is less complex than that of non-MFD systems because the positive and negative sequences are decoupled. This paper shows that if a system is incorrectly assumed to be MFD, this will lead to an erroneous or ambiguous estimation of the equivalent impedance. 
\end{abstract}




\section{Introduction}
The stability of AC power systems with a high penetration of power electronics is very difficult to analyze. The combination of multiple non-linearities and fast dynamics stemming from controllers adds significant complexity to the analysis. Impedance-based analysis of AC power systems is a relevant and practical tool in this respect because it reduces the system into a source and load subsystem, and analyses the dynamic interactions between the two subsystem equivalents \cite{Sun2009} \cite{Belkhayat1997}. The method is based on existing techniques for DC-systems, first applied in \cite{middlebrook1976}.

This method has some highly appealing properties. First, it considers the subsystems as “black-boxes”, i.e. detailed knowledge of the parameters and properties of the system is not required as long as measurements can be obtained at its terminals. Furthermore, the impedance equivalents can be extracted based on measured signals in a real system. The most accurate method for this purpose is based on frequency scanning \cite{Francis2011}\nocite{Familiant2009}-\cite{Huang2009}. However, this method requires advanced and dedicated equipment, and has limited real-time applicability. There are several alternative methods which can estimate impedance closer to real-time, and with low or zero additional hardware requirements. Examples are binary sequence injection \cite{Roinila2014}, impulse response \cite{Cespedes2014}, Kalman filtering \cite{Hoffmann2013} and recurrent neural networks \cite{Xiao2010}. However, the accuracy of these methods has not been extensively investigated in any comparative or validation studies.

When an impedance equivalent is established, it can be used for several purposes. Analytical impedance models for relatively simple systems were derived in \cite{Wen2015b}\nocite{Cespedes2014b}\nocite{Wang2014}-\cite{Turner2013}. System stability can be assessed based on these models through the Generalized Nyquist Criterion (GNC) \cite{Desoer1980}. Other stability criteria based on impedance models are described in \cite{Belkhayat1997},\cite{Liu2015}\nocite{Wildrick1995}\nocite{Sun2011}-\cite{Burgos2010}. Impedance equivalents have also been verified through experimental studies \cite{Wen2015b},\cite{Cespedes2014b},\cite{Dong2015}\nocite{Wen2015}-\cite{Wen2016}.

Previous work in this field can be grouped into two approaches. The first analyzes the system in the sequence domain using symmetric components (e.g. \cite{Sun2009},\cite{Roinila2014},\cite{Cespedes2014},\cite{bing2010thesis}), whereas the other applies the synchronous (\textit{dq}) reference frame (e.g.  \cite{Belkhayat1997},\cite{Francis2011},\cite{Wen2015b},\cite{Valdiva2012}). Both domains have certain advantages and disadvantages, but neither a rigorous comparison nor an attempt to bridge them have yet been conducted. The present paper shows mathematically how the two impedance domains are related to each other, and that they can be viewed as equivalent in terms of stability.

This paper makes two contributions. The first involves the proposed modified definition of the \textit{2x2} sequence domain impedance matrix. In this matrix, the positive and negative sequences are shifted with twice the fundamental frequency. The coupling between these two frequencies is important in power electronic systems, and is defined as \textit{the mirror frequency effect}. The equivalence between the proposed matrix and the well-established \textit{2x2} \textit{dq} domain impedance matrix is derived, and it is proven that the Generalized Nyquist Criterion (GNC) estimates is equivalent for both matrices.

The second contribution involves the definition of the \textit{Mirror Frequency Decoupled} (MFD) system, which is a sufficient condition to avoid the mirror frequency effect. It is shown how, in such systems, the impedance matrices become reduced. Furthermore, it is shown that the original definition of sequence domain impedances \cite{Sun2009} is ambiguous unless the system is MFD.

\section{Relationship between \textit{dq} and sequence domains}\label{sec:relations}

\subsection{Assumption: Sequence domain balanced systems}
This work is based on systems that are \textit{sequence domain balanced}. In other words, if a \textit{positive} sequence current at an arbitrary frequency is injected anywhere into the system, there will be no induced \textit{negative} sequence components at the same frequency. This assumption has been applied in almost all stability analyses in previous studies (e.g. \cite{Cespedes2014}, \cite{Wen2015b}, \cite{Wang2014}). Although sequence domain unbalanced systems are not treated in the present paper, the theory explained here can be extended to cover them. However, the extension of this theory may create a more complex and abstract representation of the system. 

Even in systems that are sequence-domain balanced, it will be shown later that a current injection at a given frequency can induce a voltage shifted by twice the fundamental frequency. This phenomenon is denoted the \textit{mirror frequency effect} in this paper, and will occur in most power systems. Subsystems that do not contribute to the mirror frequency effect are defined in this work as \textit{Mirror Frequency Decoupled} (MFD). This definition and its applications are presented in section \ref{sec:mirr}. 

\subsection{Sequence domain impedance extraction in previous work}
In previous work, the sequence domain impedances have been obtained in two ways. The most complete method takes into account sequence domain unbalances as defined above \cite{Cespedes2013}. The resulting impedance relation is similar to that proposed in this paper (\ref{eq:Z_pn_2}), but does not consider the mirror frequency effect defined in section \ref{sec:mirr}. Similarly, although (\ref{eq:Z_pn_2}) takes into account mirror frequency effects, it neglects sequence domain unbalances. 

The other, more simple method for obtaining sequence domain impedance assumes that the positive and negative sequence are decoupled \cite{Sun2009},\cite{Roinila2014},\cite{Cespedes2014b}. This is equivalent to the following definition, hereafter denoted the \textit{original sequence domain impedance} definition: 

\begin{align}
Z_p=\frac{V_p}{I_p}
\nonumber \\
Z_n=\frac{V_n}{I_n}
\label{eq:Zpn_orig}
\end{align}

where $Z_p$ is denoted the positive sequence impedance and $Z_n$ is denoted the negative sequence impedance. This definition takes into account neither the sequence domain unbalances nor the mirror frequency effect. Nonetheless, it can be accurate in many cases, in particular at high frequencies as will be shown by simulations in section \ref{sec:sim}.

\subsection{Harmonic current and voltage equations}
The first step in deriving the impedance relationship between \textit{dq} and the sequence domain is to relate current and voltage components to each other in the two domains. This has been done in previous studies, e.g. \cite{zmood2001}, but the purpose of this work was not related with impedance equivalents. Another study presents an impedance derivation procedure similar to the one in the present paper \cite{bing2010thesis}. The derivation was here applied to a case with diagonal \textit{dq}-impedance matrix, and is hence a special case of the relations derived in the following sections. 

First, the \textit{dq} coordinate system is defined according to the Parks transform:
\begin{equation}\label{eq:parks}
\begin{bmatrix}
v_d\\
v_q
\end{bmatrix}
= 
\sqrt{\frac{2}{3}}
\begin{bmatrix}
\cos(\theta)  & \cos(\theta-\frac{2\pi}{3})  &  \cos(\theta+\frac{2\pi}{3}) \\
-\sin(\theta)  & -\sin(\theta-\frac{2\pi}{3})  &  -\sin(\theta+\frac{2\pi}{3}) \\
\end{bmatrix}
\begin{bmatrix}
v_a\\
v_b\\
v_c
\end{bmatrix}
 \end{equation}

For a given three-phase set of signals, $v_a$, $v_b$, $v_c$, the corresponding \textit{dq}-domain signals are given by (\ref{eq:parks}). $\theta$ is the transformation angle typically obtained from a Phase Lock Loop (PLL) or from an oscillator. In steady-state, $\theta=\omega_1 t$, where $\omega_1$ is the fundamental frequency. (\ref{eq:parks}) is here expressed in the time-domain, but is valid in the frequency-domain as well. From now on, all equations are expressed in the frequency domain. The relationship between a time domain waveform and its frequency domain components is

\begin{equation}\label{eq:time_to_freq}
v(t)=\sum\limits_{k = 0}^\infty{{v_k}\cos(\omega_k t + {\phi_k})}
\end{equation}

where $v_k$ is the amplitude of the frequency component at frequency $\omega_k$, and $\phi_k$ is the corresponding phase angle. The complex number
\begin{equation}
{V_k} = {v_k}{e^{j{\phi_k}}}
\end{equation}

is defined as the harmonic phasor at frequency $\omega_k$. From this point, the subscript \textit{k} is omitted, and the relevant frequency will be indicated.

The sequence domain, also denoted the symmetric component domain, is widely applied in power system analysis because it can decompose unbalanced three-phase systems into three balanced and decoupled subsystems:
\begin{itemize}
	\item A positive sequence subsystem (\textit{p})
	\item A negative sequence subsystem (\textit{n})
	\item A zero sequence subsystem (\textit{0})
\end{itemize}

The effect of zero sequence components is disregarded in this paper. This is the equivalent of assuming a system without a neutral wire, as well as neglecting transmission line capacitance. Note that in most applications of the sequence domain, only the fundamental frequency components are considered, whereas in this work the phasors can be related to any frequency. \textit{abc}-phasors can be related to sequence domain phasors by the symmetric component transform:

\begin{equation}\label{eq:pn_abc}
\begin{bmatrix}
V_p\\
V_n
\end{bmatrix}
= 
\begin{bmatrix}
1 & a & a^2 \\
1 & a^2 & a \\
\end{bmatrix}
\begin{bmatrix}
V_a \\
V_b \\
V_c 
\end{bmatrix}
 \end{equation}

where $V_p$ is the \textit{positive sequence} voltage phasor, and $V_n$ is the \textit{negative sequence} voltage phasor at an arbitrary frequency. $a=e^{j\frac{2\pi}{3}}$ is the complex number corresponding to a $120^o$ phase shift. It can be shown that sequence domain phasors are related to \textit{dq}-domain phasors as follows:

\begin{align}
\begin{bmatrix}
V_p\\
V_n
\end{bmatrix}
&= 
\frac{1}{\sqrt{6}}
\begin{bmatrix}
V_d+jV_q \\
0\\
\end{bmatrix}
,\omega_p=\omega_{dq}+\omega_1 \nonumber
\\ 
\begin{bmatrix}
V_p\\
V_n
\end{bmatrix}
&= 
\frac{1}{\sqrt{6}}
\begin{bmatrix}
0 \\
V_d-jV_q\\
\end{bmatrix}
,\omega_n=\omega_{dq}-\omega_1
\label{eq:dq_to_pn}
\end{align}

In other words, a general \textit{dq} voltage phasor at frequency $\omega_{dq}$ is equivalent to two sequence domain voltage phasors at different frequencies, shifted by the fundamental:
\begin{itemize}
\item Positive sequence voltage at $\omega_p=\omega_{dq}+\omega_1$
\item Negative sequence voltage at $\omega_n=\omega_{dq}-\omega_1 $
\end{itemize}

Similarly, it can be shown that for a given positive or negative sequence voltage phasor, the corresponding \textit{dq}-domain phasors are:

\begin{align}
\begin{bmatrix}
V_d\\
V_q
\end{bmatrix}
&= 
\sqrt{\frac{3}{2}}
V_p
\begin{bmatrix}
1 \\
-j\\
\end{bmatrix}
,\omega_{dq}=\omega_{p}-\omega_1 \nonumber
\\ 
\begin{bmatrix}
V_d\\
V_q
\end{bmatrix}
&= 
\sqrt{\frac{3}{2}}
V_n
\begin{bmatrix}
1 \\
j\\
\end{bmatrix}
,\omega_{dq}=\omega_{n}+\omega_1
\label{eq:pn_to_dq}
\end{align}

By definition, (\ref{eq:parks})-(\ref{eq:pn_to_dq}) also applies to currents.

\subsection{Illustration of harmonic phasor relations}
\begin{figure}[ht!]
     \centering
     \includegraphics[width=0.48\textwidth]{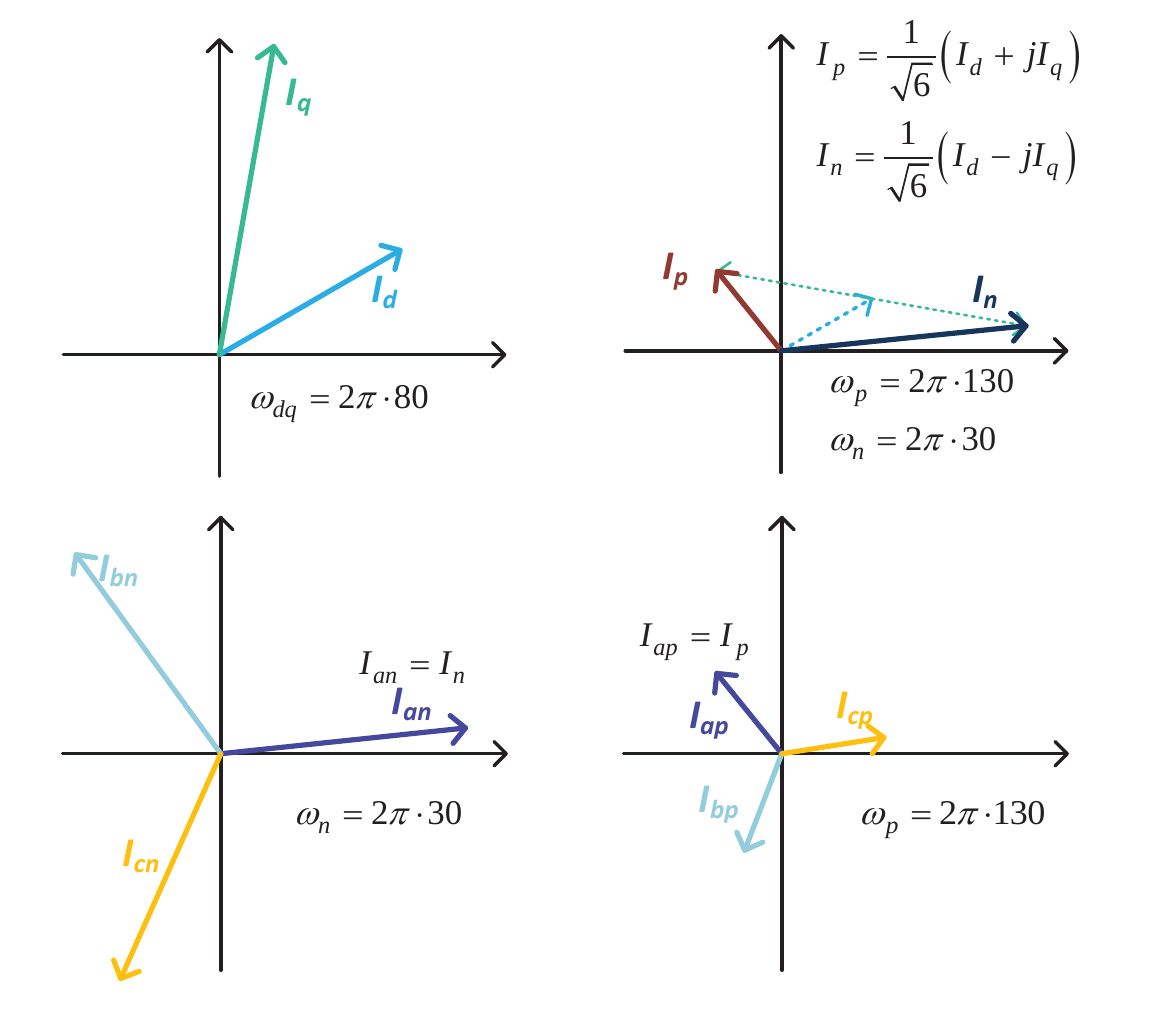}
     \caption{Harmonic phasors corresponding to (\ref{eq:phasor_example}). The \textit{80 Hz}-waveform at $\omega_{dq}$ equals the sum of a positive sequence waveform at $\omega_p=\omega_{dq}+\omega_1$ and a negative sequence waveform at $\omega_n=\omega_{dq}-\omega_1$}
     \label{fig:illustration_phasor}
\end{figure}

Equation (\ref{eq:dq_to_pn}) is illustrated in Figure \ref{fig:illustration_phasor}. A perturbation in a \textit{dq}-domain current waveform at a randomly selected frequency of \textit{80 Hz} is modelled as follows:

\begin{equation}\label{eq:phasor_example}
\begin{rcases}
I_{d}=0.06\angle 80^o\\
I_{q}=0.09\angle 30^o
\end{rcases},
\omega_{dq}=2\pi\cdot 80
\end{equation}

The waveform is expressed in the abc-domain using the inverse of (\ref{eq:parks}). FFT is used to calculate the frequency domain harmonic phasors in both domains. These are drawn in the complex plane in Figure \ref{fig:illustration_phasor}. It is seen that the single \textit{dq}-tone at \textit{80 Hz} is transformed into a \textit{30 Hz} and a \textit{130 Hz} component in the \textit{abc}-domain. Moreover, it is seen that the \textit{30 Hz} waveforms are pure negative sequence, whereas the \textit{130 Hz} waveforms are pure positive sequence, which is consistent with (\ref{eq:dq_to_pn}). The sequence domain phasors are calculated from the \textit{dq}-domain phasors by this equation. Note also that the sequence domain phasors satisfies the following equation for balanced \textit{abc}-components: $I_{ap}=I_p$ and $I_{an}=I_n$.

\begin{figure*}[ht]
     \centering
     \includegraphics[width=0.95\textwidth]{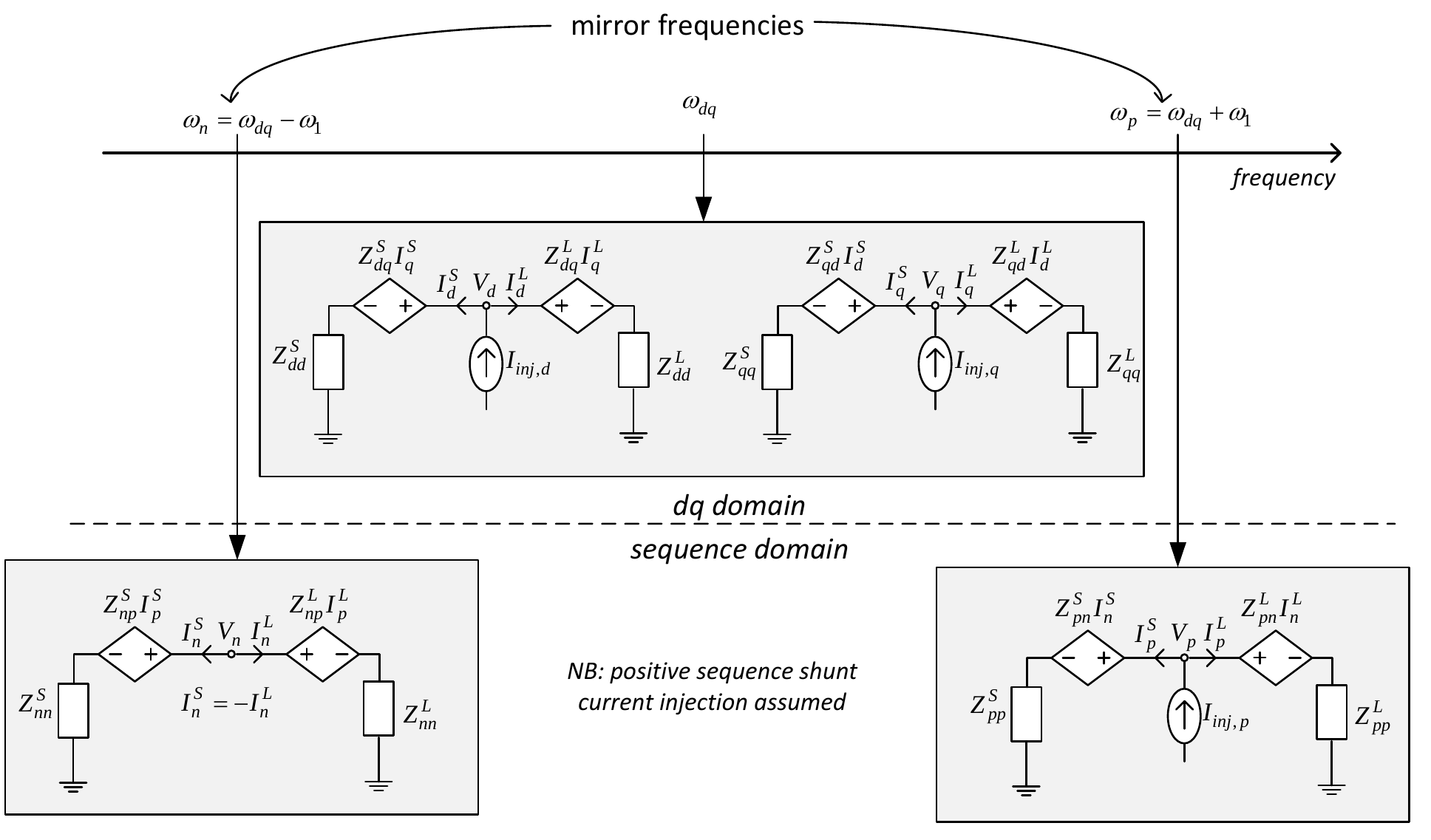}
     \caption{Circuit equivalents of impedance equations in the \textit{dq} and sequence domain. Mirror frequency coupling illustrated by voltage sources in the sequence domain.}
     \label{fig:MFE_illustration}
\end{figure*}

\section{Modified sequence domain impedance definition}\label{sec:Zpn}
In the previous section, current and voltage phasors were considered separately. This section will relate them to each other through impedance. The main contribution of the paper is then outlined, which is a \textit{2x2} impedance matrix composed of positive and negative sequence impedances at two different frequencies. The proposed \textit{2x2} matrix has many appealing properties, as will be highlighted in further sections.

To provide a better overview, all current, voltages and impedances are summarized in Table \ref{tab:frequency_overview}. These definitions highlight a key point in the paper: positive and negative sequences are defined at two different frequencies, shifted by twice the fundamental frequency. The two frequencies are also denoted \textit{mirror frequencies} later in the paper.

\begin{table}
\centering
\caption{Overview of currents, voltages, and impedances and the corresponding frequencies at which they are defined}
\begin{tabular}{c|c}
\textbf{Parameter} & \textbf{Frequency} \\ \hline
$V_d$ / $I_d$ & $\omega_{dq}$ \\
$V_q$ / $I_q$ & $\omega_{dq}$ \\
$\bm{Z_{dq}}$ & $\omega_{dq}$ \\
$V_p$ / $I_p$ & $\omega_{p}=\omega_{dq}+\omega_1$ \\
$V_n$ / $I_n$ & $\omega_{n}=\omega_{dq}-\omega_1$ \\
$Z_{pp}$ & $\omega_p$ \\
$Z_{nn}$ & $\omega_n$ \\
$Z_{pn}$ & $\omega_n \rightarrow \omega_p$ \\
$Z_{np}$ & $\omega_p \rightarrow \omega_n$  \\
$Z_{p}$ & $\omega_p$  \\
$Z_{n}$ & $\omega_n$  
\end{tabular}
\label{tab:frequency_overview}
\end{table}

\subsection{Definition and fundamental equations}
The basis for the following derivations is the generalized Ohms law in the \textit{dq}-domain:

\begin{equation}\label{eq:Ohms_dq}
\begin{bmatrix}
V_d \\
V_q
\end{bmatrix}
=
\begin{bmatrix}
Z_{dd} & Z_{dq} \\
Z_{qd} & Z_{qq}
\end{bmatrix}
\begin{bmatrix}
I_d \\
I_q
\end{bmatrix}
=
\bm{Z_{dq}}
\begin{bmatrix}
I_d \\
I_q
\end{bmatrix}
\end{equation}

where $\bm{Z_{dq}}$ is a \textit{2x2} matrix of complex numbers as a function of frequency. This equation will now be transformed into the sequence domain. 

As shown by (\ref{eq:pn_to_dq}), any set of \textit{dq}-domain phasors can be written as the sum of positive sequence phasors at $\omega_{dq}+\omega_1$ and negative sequence phasors at $\omega_{dq}-\omega_1$. Substituting from (\ref{eq:pn_to_dq}) gives a modified version of (\ref{eq:Ohms_dq}):

\begin{equation}\label{eq:Ohms_dq_mod}
V_p
\begin{bmatrix}
1 \\
-j
\end{bmatrix}
+
V_n
\begin{bmatrix}
1 \\
j
\end{bmatrix}
=
\bm{Z_{dq}}
\left(
I_p
\begin{bmatrix}
1 \\
-j
\end{bmatrix}
+
I_n
\begin{bmatrix}
1 \\
j
\end{bmatrix}
\right)
\end{equation}

The definitions shown in Table \ref{tab:frequency_overview} are based on this convenient representation. The next step is to multiply (\ref{eq:Ohms_dq_mod}) with $\frac{1}{2}\begin{bmatrix}1 & j\end{bmatrix}$ and $\frac{1}{2}\begin{bmatrix}1 & -j\end{bmatrix}$, leading to the following two equations:

\begin{align}
V_p
=
I_p
\left(
\frac{1}{2}
\begin{bmatrix}1 & j\end{bmatrix}
\bm{Z_{dq}}
\begin{bmatrix}1\\-j\end{bmatrix}
\right)
&+
I_n
\left(
\frac{1}{2}
\begin{bmatrix}1 & j\end{bmatrix}
\bm{Z_{dq}}
\begin{bmatrix}1\\j\end{bmatrix}
\right)
\nonumber \\
V_n
=
I_p
\left(
\frac{1}{2}
\begin{bmatrix}1 & -j\end{bmatrix}
\bm{Z_{dq}}
\begin{bmatrix}1\\-j\end{bmatrix}
\right)
&+
I_n
\left(
\frac{1}{2}
\begin{bmatrix}1 & -j\end{bmatrix}
\bm{Z_{dq}}
\begin{bmatrix}1\\j\end{bmatrix}
\right)
\label{eq:Ohms_pn}
\end{align}
 
 (\ref{eq:Ohms_pn}) can conveniently be rewritten with matrix notation by defining the \textit{modified sequence domain impedance matrix}:

\begin{equation}\label{Ohms_pn_2}
\begin{bmatrix}
V_p \\
V_n
\end{bmatrix}
=
\begin{bmatrix}
Z_{pp} & Z_{pn} \\
Z_{np} & Z_{nn}
\end{bmatrix}
\begin{bmatrix}
I_p \\
I_n
\end{bmatrix}
=
\bm{Z_{pn}}
\begin{bmatrix}
I_p \\
I_n
\end{bmatrix}
\end{equation}

with the four impedances substituted from (\ref{eq:Ohms_pn}):

\begin{equation}\label{eq:Z_pn}
\bm{Z_{pn}}
=
\frac{1}{2}
\begin{bmatrix}
	\begin{bmatrix}1 & j\end{bmatrix}
		\bm{Z_{dq}}
	\begin{bmatrix}1 \\ -j\end{bmatrix}
&
	\begin{bmatrix}1 & j\end{bmatrix}
		\bm{Z_{dq}}
	\begin{bmatrix}1 \\ j\end{bmatrix}
\\			
	\begin{bmatrix}1 & -j\end{bmatrix}
		\bm{Z_{dq}}
	\begin{bmatrix}1 \\ -j\end{bmatrix}
&
	\begin{bmatrix}1 & -j\end{bmatrix}
		\bm{Z_{dq}}
	\begin{bmatrix}1 \\ j\end{bmatrix}
\end{bmatrix}
\end{equation}

The impedances have the following physical interpretation:
\begin{itemize}
\item $Z_{pp}$: Measures the positive sequence voltage at $\omega_p$ induced by a positive sequence current at $\omega_p$
\item $Z_{pn}$: Measures the positive sequence voltage at $\omega_p$ induced by negative sequence current at $\omega_n$
\item $Z_{np}$: Measures the negative sequence voltage at $\omega_n$ induced by positive sequence current at $\omega_p$
\item $Z_{nn}$: Measures the negative sequence voltage at $\omega_n$ induced by a negative sequence current at $\omega_n$
\end{itemize}

(\ref{eq:Z_pn}) can be rewritten in a more compact form as a linear transformation by the unitary matrix $A_Z$:

\begin{align}
\bm{Z_{pn}}
&=
A_Z \cdot \bm{Z_{dq}} \cdot A_Z^{-1}
\nonumber \\
\bm{Z_{dq}}
&=
A_Z^{-1} \cdot \bm{Z_{pn}} \cdot A_Z
\nonumber \\
A_Z=\frac{1}{\sqrt{2}}
\begin{bmatrix}1 & j\\1 & -j\end{bmatrix}
&,
A_Z^{-1}=A_Z^*
=\frac{1}{\sqrt{2}}
\begin{bmatrix}1 & 1\\-j & j\end{bmatrix}
\label{eq:Z_pn_2}
\end{align}

where * denotes complex conjugate transpose. 

The corresponding admittance equations can be obtained in the same way by interchanging voltages with currents in the derivation process:
\begin{align}
\bm{Y_{pn}}
&=
A_Z \cdot \bm{Y_{dq}} \cdot A_Z^{-1}
\nonumber \\
\bm{Y_{dq}}
&=
A_Z^{-1} \cdot \bm{Y_{pn}} \cdot A_Z
\end{align}

The generalized Ohms law and the mirror frequency effect are illustrated as circuit equivalents in Figure \ref{fig:MFE_illustration}. The figure assumes positive sequence shunt current injection at frequency $\omega_p$, but corresponding equivalents can be made for other injection choices. In the \textit{dq}-domain, all signals are expressed using the same frequency $\omega_{dq}$. The cross coupling between the \textit{d}- and \textit{q}-axis is represented by current dependent voltage sources. 

\subsection{Positive sequence impedances below the fundamental frequency}
\label{subsec:pos_bel_fund}
In this paper, the modified sequence domain impedance definition has not thus far been defined for positive sequence at frequencies below the fundamental frequency. Given that $\omega_p=\omega_{dq}+\omega_1$, the \textit{dq}-domain frequency $\omega_{dq}$ will be negative for $0\leq\omega_p<\omega_1$. To extend the impedance definition to positive sequence below the fundamental frequency, it is first important to note that a balanced three-phase signal with a negative frequency is equivalent to a balanced three-phase signal with positive frequency at the same absolute value. Only the phase order needs to be changed, i.e. a positive sequence signal becomes a negative sequence, and vice versa.

Secondly, note that when $0\leq \omega_{dq} <\omega_1$, the negative sequence frequency $\omega_n=\omega_{dq}-\omega_1$ is negative. Consequently, the negative sequence impedance $Z_{nn}$ is associated with a negative frequency. Based on the discussion above, the negative sequence impedance at a negative frequency is equal to the positive sequence impedance at a positive frequency with the same absolute value. Thus, a positive sequence impedance below the fundamental frequency can be defined. 

To summarize, an example of \textit{dq}-domain and corresponding sequence domain frequencies are given in Table \ref{tab:frequency_examples}.

\begin{table}
\centering
\caption{Example of \textit{dq}- and sequence domain frequencies}
\begin{tabular}{c|c |c}
\textbf{$\omega_{dq}$} & $\omega_{p}$ & $\omega_{n}$ \\ \hline
500 & 550 & 450 \\
120 & 170 & 70 \\
65 & 115 & 15 \\
40 & 90 & -10$\rightarrow$10$^*$ \\
15 & 65 & -35$\rightarrow$35$^*$
\end{tabular}
\\$^*$ negative sequence frequencies where the impedances are transformed to positive sequence as explained in section \ref{subsec:pos_bel_fund}
\label{tab:frequency_examples}
\end{table}

\subsection{Nyquist stability criterion equivalence}\label{subsec:GNC}
The Generalized Nyquist stability Criterion (GNC) has been widely applied in previous studies to analyze the stability of power electronics systems (e.g. \cite{Wang2014},\cite{Sun2011},\cite{Wen2015}). The criterion is mathematically formulated in \cite{Desoer1980}, and was first applied to AC power electronic systems in \cite{Belkhayat1997}. It will now be shown that when the GNC is applied to the \textit{dq} and the modified sequence domains, the results are identical.

Assuming the \textit{dq}-domain, the basis for the stability criterion is the system transfer function between source and load:
\begin{equation}
\bm{h}=\text{inv}(\bm{I + Z_{dq}^S Y_{dq}^L})
\end{equation}

For convenience, the minor-loop gain $\bm{L_{dq}}$ is defined as:
\begin{equation}
\bm{L_{dq}} = \bm{Z_{dq}^S Y_{dq}^L}
\end{equation}

The eigenvalues of $\bm{L_{dq}}$ can be found by solving:
\begin{equation}
\text{det} \left( \bm{L_{dq}} + \lambda_{dq}\bm{I} \right) = 0
\end{equation}

Assume in the following that the source is stable when connected to an ideal load, and that the load is stable when connected to an ideal source. The GNC then states that the system is stable if and only if the characteristic loci of $\bm{L_{dq}}$ do not encircle the point $(-1,0)$ when drawn in the complex plane. 

The minor-loop gain in the sequence domain can be expressed as:
\begin{align}
\bm{L_{pn}}&=\bm{Z_{pn}^S Y_{pn}^L}
=
\left( A_Z \cdot \bm{Z_{dq}^S} \cdot A_Z^{-1} \right)
\left( A_Z \cdot \bm{Y_{dq}^L} \cdot A_Z^{-1} \right)
\nonumber \\
&=
A_Z \cdot \bm{Z_{dq}^S Y_{dq}^L} \cdot A_Z^{-1}
=
A_Z \cdot \bm{L_{dq}} \cdot A_Z^{-1}
\end{align}

The following equations prove that $\lambda_{dq}$, the eigenvalues of $\bm{L_{dq}}$, are equal to $\lambda_{pn}$, the eigenvalues of $\bm{L_{pn}}$.

\begin{align}
0 &=\text{det}\left( \bm{L_{pn}}-\lambda_{pn}\bm{I}\right)
= \text{det}\left( A_Z \cdot \bm{L_{dq}} \cdot A_Z^{-1} -\lambda_{pn}\bm{I}\right)
\nonumber \\
&=
\text{det}\left( A_Z \cdot \bm{L_{dq}} \cdot A_Z^{-1} 
-A_Z \cdot \left(\lambda_{pn} \bm{I}\right) \cdot A_Z^{-1}\right)
\nonumber \\
&=
\text{det}(A_Z)\cdot \text{det}\left( \bm{L_{dq}} - \lambda_{pn}\bm{I}\right) \cdot \text{det}\left(A_Z^{-1} \right)
\nonumber \\
&=
\text{det} \left( \bm{L_{dq}} - \lambda_{pn}\bm{I} \right) = 0
\nonumber \\
& \Rightarrow \lambda_{pn}=\lambda_{dq}
\label{eq:GNCproof}
\end{align}

Consequently, the stability analysis by GNC gives identical results in the \textit{dq}- and sequence domains when the \textit{modified} definition (\ref{eq:Z_pn_2}) is applied. By contrast, if the \textit{original} definition (\ref{eq:Zpn_orig}) is used, and one or both subsystems are \textit{not} mirror frequency decoupled (see section \ref{sec:mirr}), the calculated stability is different from the \textit{dq}-domain stability calculations.

\section{The mirror frequency effect}\label{sec:mirr}
The term \textit{mirror frequency effect} is defined in this paper to provide further insight into the properties of impedance models. When subjected to a harmonic disturbance at a given frequency, power electronic converters will respond with induced current/voltages at the same frequency, but also at the so-called mirror frequency. The mirror frequency is shifted with twice the fundamental frequency, and the direction depends on whether the disturbance is a positive or negative sequence. This effect complicates the analysis of the system because it violates the assumption of linearity. However, when a system is assumed to be \textit{sequence domain balanced} as discussed in section \ref{sec:relations}, linearity can be regained by either
\begin{itemize}
	\item applying the $dq$-domain
	\item applying the \textit{modified sequence domain} defined by (\ref{eq:Z_pn_2})
\end{itemize}

Of note, sequence domain balanced systems may still contain the mirror frequency effect as these two properties are independent. The remainder of this section will focus on sequence domain balanced systems lacking the mirror frequency effect. Such systems are hereafter denoted \textit{Mirror Frequency Decoupled (MFD)}.

\subsection{Mirror Frequency Decoupled (MFD) systems}
A subsystem is said to be \textit{Mirror Frequency Decoupled} (MFD) if, when subjected to a harmonic disturbance at an arbitrary frequency, it only responds with current/voltages at the same frequency. With reference to Figure \ref{fig:MFE_illustration}, this is equivalent to removing the current-dependent voltage sources in the sequence domain. In other words, $Z_{pn}=Z_{np}=0$. MFD systems have several interesting properties presented below.

\subsection{Impedance matrices of MFD systems}

MFD subsystems have sequence- and \textit{dq}-domain impedance matrices of the following form:

\begin{align}
\bm{Z_{pn}}\Big|_{MFD}
&=
\begin{bmatrix} Z_{pp} & 0 \\ 0 & Z_{nn} \end{bmatrix}
\nonumber \\
\bm{Z_{dq}}\Big|_{MFD}
&=
\begin{bmatrix} Z_{x} & Z_{y} \\ -Z_{y} & Z_{x} \end{bmatrix}
\label{eq:Z_mfd}
\end{align}

where $Z_{pn}=Z_{np}=0$ by the definition in the previous section. The \textit{dq}-domain matrix is skew symmetric with $Z_{dd}=Z_{qq}=Z_x$ and $Z_{dq}=-Z_{qd}=Z_y$. The proof of these relations is presented in Appendix \ref{app:mfd}. In a previous study a similar result was found under the assumption of $Z_{dq}=Z_{qd}=0$ \cite{bing2010thesis}. 

Since $Z_{pn}$ is diagonal, the original impedance definition (\ref{eq:Zpn_orig}) is equivalent to the modified (\ref{eq:Z_pn_2}) for MFD-systems.

\subsection{\textit{dq} impedance extraction in MFD systems}
It has been argued that sequence domain impedances are easier to obtain than \textit{dq}-domain impedances due to the decoupling between positive and negative sequence \cite{Sun2009},\cite{Cespedes2014b}. Furthermore, the sequence domain impedance can be obtained from a single measurement with no need for matrix inversions because of this decoupling. However, it has been shown in the previous section that sequence domain impedances can been assumed to be decoupled only if the subsystem is MFD. In contrast with the statements of previous work, the following equations show that \textit{dq}-domain impedances can also be obtained from a single measurement in this case. Combining (\ref{eq:Ohms_dq}) with (\ref{eq:Z_mfd}) gives:

\begin{align}
Z_x=Z_{dd}=Z_{qq}=\frac{V_d I_d + V_q I_q}{I_d^2 + I_q^2}
\nonumber \\
Z_y=Z_{dq}=-Z_{qd}=\frac{V_d I_q - V_q I_d}{I_d^2 + I_q^2}
\end{align}

Consequently, only a single measurement is needed to obtain the \textit{dq} impedance matrix in a MFD subsystem.

\subsection{Sources to mirror frequency coupling}
Mirror frequency coupling is introduced in all parts of the power system where (\ref{eq:Z_mfd}) is not satisfied. For example:
\begin{itemize}
\item Phase-Lock-Loops (PLL)
\item Converter current controllers with unequal structure and/or parameter values in the \textit{d-} and \textit{q}-axis
\item DC-link voltage control systems
\item Active and reactive power controllers
\item Salient-pole synchronous machines
\end{itemize}

The analytical impedance calculation in the \textit{dq}-domain is described in e.g. \cite{Wen2015b}, where the coupling related to the first four bullet-points can be identified. Note that all transfer functions must be identical in \textit{d}- and \textit{q}-axis in order for the subsystem to be MFD. Furthermore, all cross-coupling between \textit{d}- and \textit{q}-axis must have opposite sign. The synchronous machine is also mentioned because it is a vital part of many power systems, and it possesses mirror frequency coupling if the reluctance in \textit{d}- and \textit{q}-axes differ. Although mirror frequency coupling is independent of power electronics, it is clear from the bullet points above that power electronics systems introduces many instances for this to occur. 

\section{Validation by numeric simulation}\label{sec:sim}
\subsection{Obtaining impedances through simulation}
\begin{figure}[ht!]
     \centering
     \includegraphics[width=0.38\textwidth]{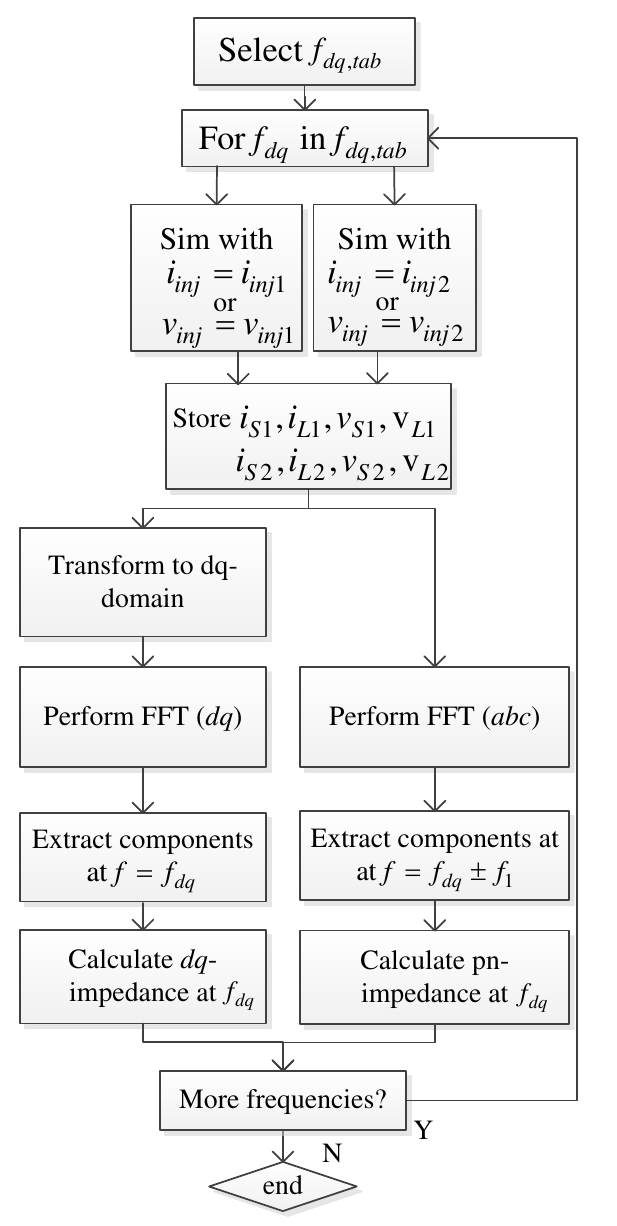}
     \caption{Illustration of simulation method to obtain both \textit{dq} and sequence domain impedances as a function of frequency. The flowchart is valid for both shunt current and series voltage injection.}
     \label{fig:flowchart}
\end{figure}

The method for obtaining impedances through simulation is best explained by the flowchart in Figure \ref{fig:flowchart}. This method is able to calculate both the \textit{dq}-domain and sequence domain impedances in an integrated process. The first step is to select a vector of frequencies $f_{dq,tab}$, i.e. the frequencies at which the impedances shall be calculated. Note that these frequencies are expressed in the \textit{dq}-domain.

\begin{figure}[ht!]
     \centering
     \includegraphics[width=0.49\textwidth]{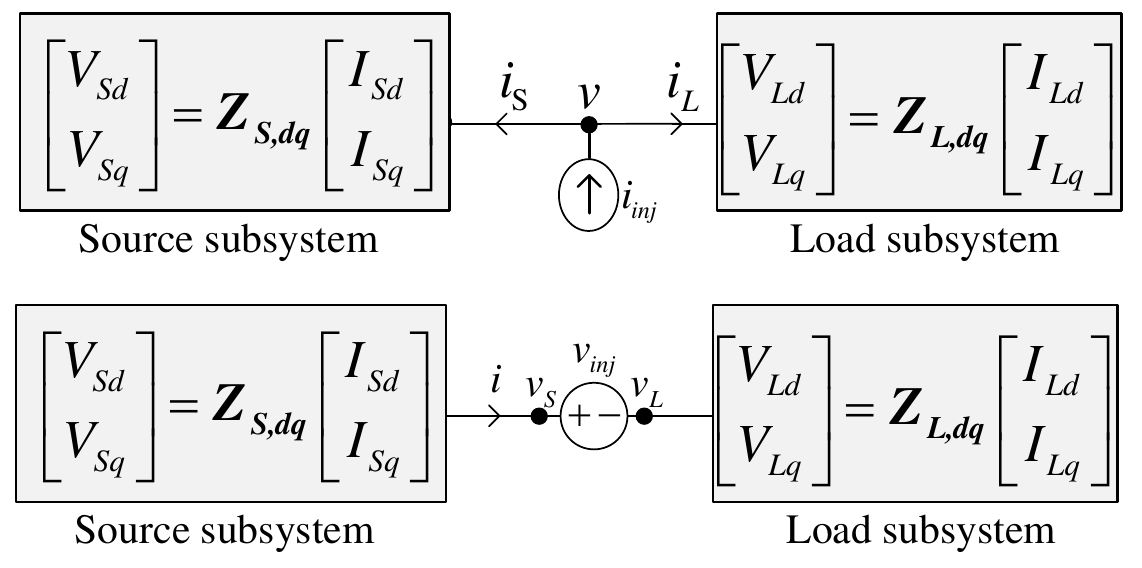}
     \caption{Illustration of the two injection methods (shunt and series) for a general source and load subsystem.}
     \label{fig:injection}
\end{figure}

The system can be simulated under \textit{either} shunt current or series voltage injection. The difference between these two methods is illustrated in Figure \ref{fig:injection}. If shunt current is used, the following three-phase perturbation signals will be injected:
\begin{align}
i_{inj1} &= I_{inj}
\begin{bmatrix}
\sin\left( \left[\omega_{inj}+\omega_1\right]t\right)
\\
\sin\left( \left[\omega_{inj}+\omega_1\right]t - \frac{2\pi}{3} \right)
\\
\sin\left( \left[\omega_{inj}+\omega_1\right]t + \frac{2\pi}{3} \right)
\end{bmatrix}
\nonumber \\
i_{inj2} &= I_{inj}
\begin{bmatrix}
\sin\left( \left[\omega_{inj}-\omega_1\right]t\right)
\\
\sin\left( \left[\omega_{inj}-\omega_1\right]t + \frac{2\pi}{3} \right)
\\
\sin\left( \left[\omega_{inj}-\omega_1\right]t - \frac{2\pi}{3} \right)
\end{bmatrix}
\end{align}

 If series voltage injection is applied, $i$ can be replaced with $v$. The two sets of signals need to have different frequencies because linear independent injections are required when solving for the impedance matrices  (\ref{eq:solve_Zdq})-(\ref{eq:solve_Zpn}). The selection of injection signals is discussed in \cite{Francis2011}.

The needed output from simulations are the current and voltage signals shown in Figure \ref{fig:injection}. Note that the injection signal itself is not needed in impedance calculations. After converting time-domain signals to the frequency domain as described in the flowchart, the following equations can be used to find the impedances in the two domains:

\begin{align}
&\begin{bmatrix} Z_{dd} & Z_{dq}\\ Z_{qd} & Z_{qq} \end{bmatrix}
=
\begin{bmatrix} V_{d1} & V_{d2}\\ V_{q1} & V_{q2} \end{bmatrix}
\begin{bmatrix} I_{d1} & I_{d2}\\ I_{q1} & I_{q2} \end{bmatrix}^{-1}
\label{eq:solve_Zdq}\\
&\begin{bmatrix} Z_{pp} & Z_{pn}\\ Z_{np} & Z_{nn} \end{bmatrix}
=
\begin{bmatrix} V_{p1} & V_{p2}\\ V_{n1} & V_{n2} \end{bmatrix}
\begin{bmatrix} I_{p1} & I_{p2}\\ I_{n1} & I_{n2} \end{bmatrix}^{-1}
\label{eq:solve_Zpn}
\end{align}

After these two matrices are established, all other impedance expressions in the paper can be derived based on them.

\subsection{Case study description}
\begin{figure}[ht!]
     \centering
     \includegraphics[width=0.49\textwidth]{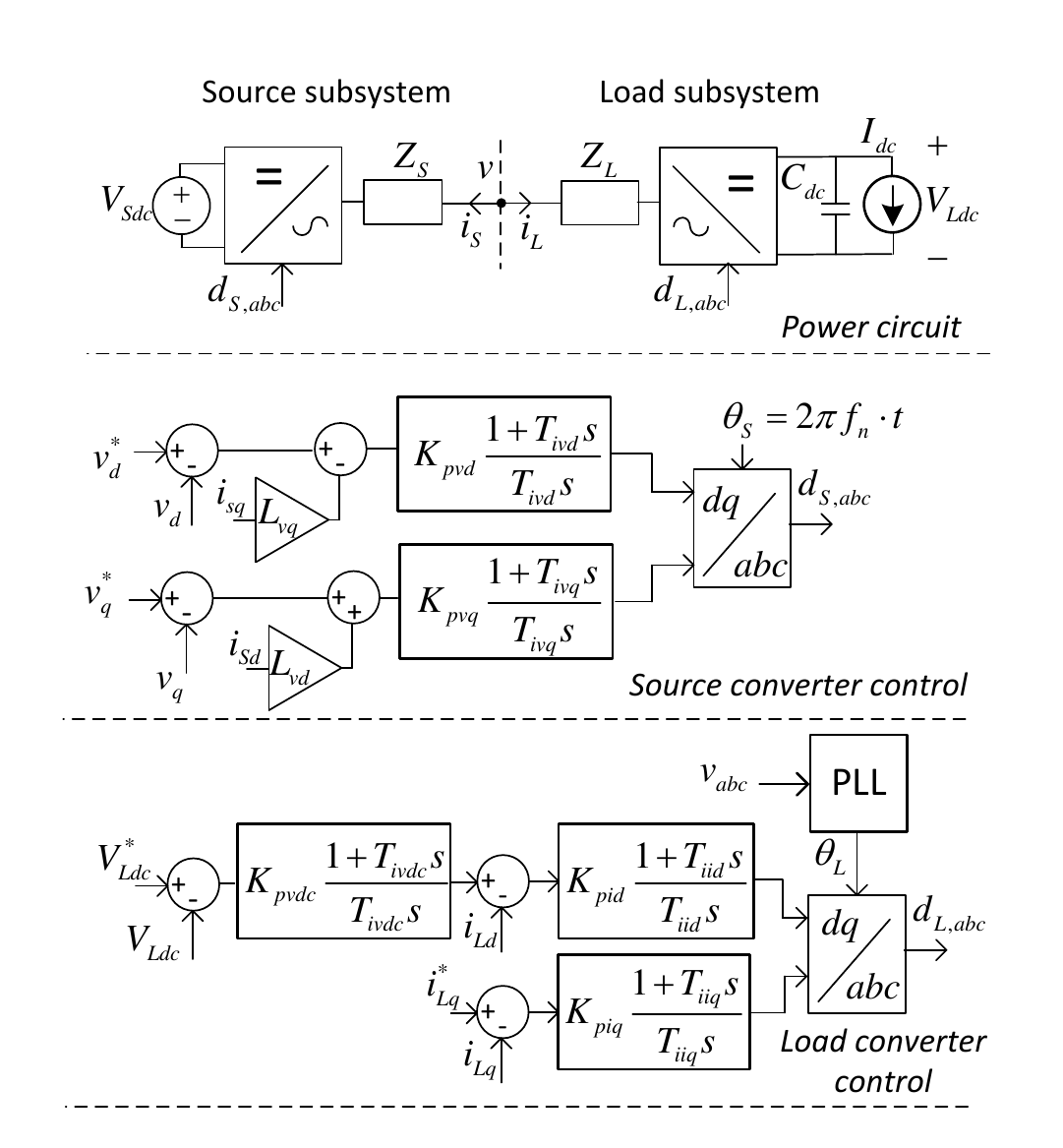}
     \caption{Detailed schematic of Case A}
     \label{fig:CaseA}
\end{figure}

\begin{figure}[ht!]
     \centering
     \includegraphics[width=0.47\textwidth]{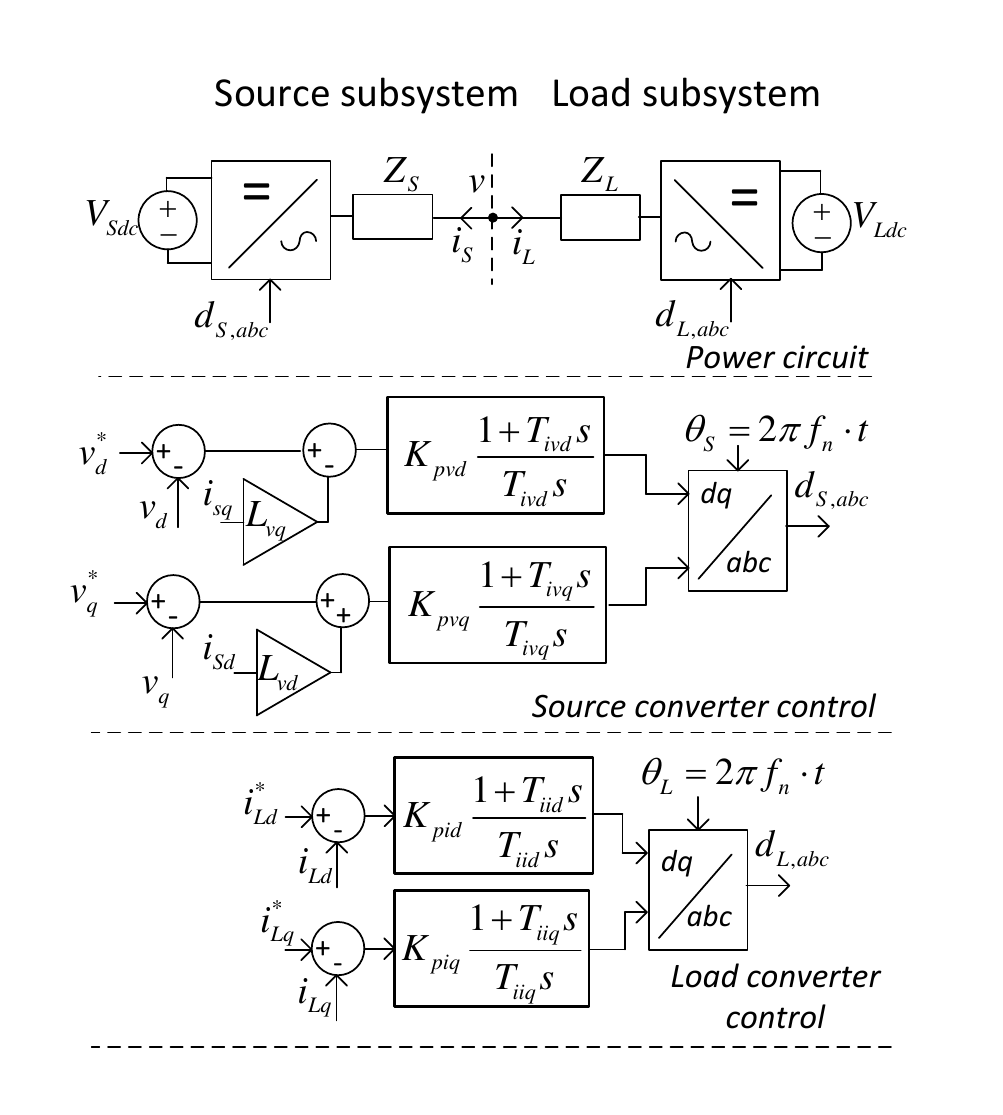}
     \caption{Detailed schematic of Case B}
     \label{fig:CaseB}
\end{figure}

In this section, the validity of the previously derived expressions is checked through numeric simulations. Simulation cases A and B are developed in MATLAB/Simulink, see Figure \ref{fig:CaseA} and Figure \ref{fig:CaseB}. Both cases consist of a source converter and a load converter. The control systems operates in the \textit{dq}-domain.

In Case A the source converter controls the voltage $v$ according to set-points and the virtual inductances $L_{vd}$ and $L_{vq}$. The converter is synchronized to the fixed clock signal $\theta_S = 2\pi f_n\cdot t$. The load converter operates with DC-voltage control and reactive current control, and a current source $I_{dc}$ consumes power at the DC-side. The converter is synchronized to the grid by a Phase Lock Loop (PLL).

\newpage

The following sources to mirror frequency coupling is present in Case A \textit{source subsystem}:

\begin{itemize}
    \item $L_{vd}\neq L_{vq}$
    \item $K_{pvd}\neq K_{pvq}$
    \item $T_{ivd}\neq T_{ivq}$
\end{itemize}

In Case A \textit{load subsystem}, the following sources to mirror frequency coupling is present:
    
\begin{itemize}    
    \item DC-link voltage controller
    \item PLL is connected to a non-stiff point in the grid
    \item $K_{pid}\neq K_{piq}$
    \item $T_{pid}\neq T_{piq}$
\end{itemize}

To better illustrate the findings in the paper, Case A has been divided into two subcases, \textbf{Case A1} and \textbf{Case A2}. In Case A1 both subsystems are mirror frequency coupled according to the bullet-lists above. However, in Case A2 all mirror frequency couplings in the source subsystem are removed by setting $L_{vd}=L_{vq}$, $K_{pvd}=K_{pvq}$ and $T_{ivd}= T_{ivq}$. Hence, in Case A2 the source subsystem is MFD, while the load subsystem is not. The differences between these two cases will be discussed in the result section.

In Case B all mirror frequency coupling is removed in both subsystems, leading to a complete MFD system. The source subsystem is identical to the one in Case A2. In the load subsystem the converter is connected to a constant voltage at the DC-side, which eliminates the need for DC voltage control. The control system consists of current controllers with set-points ${i_{Ld}^*}$ and ${i_{Ld}^*}$. The PI-controller parameters are identical in \textit{d}- and \textit{q}-axis. Furthermore, the converter does not contain a PLL, but is instead synchronized to the fixed ramp $\theta_L = 2\pi f_n\cdot t$ in the same way as the source converter. Consequently, all mirror frequency coupling sources from the above bullet-list have been removed.

Parameter values applied in the simulation cases are given in Appendix \ref{app:params}.

\subsection{Simulation results - $\bm{Z_{dq}}$ and $\bm{Z_{pn}}$}
\label{subsec:sim1}
\begin{figure}[ht!]
     \centering
     \includegraphics[width=0.48\textwidth]{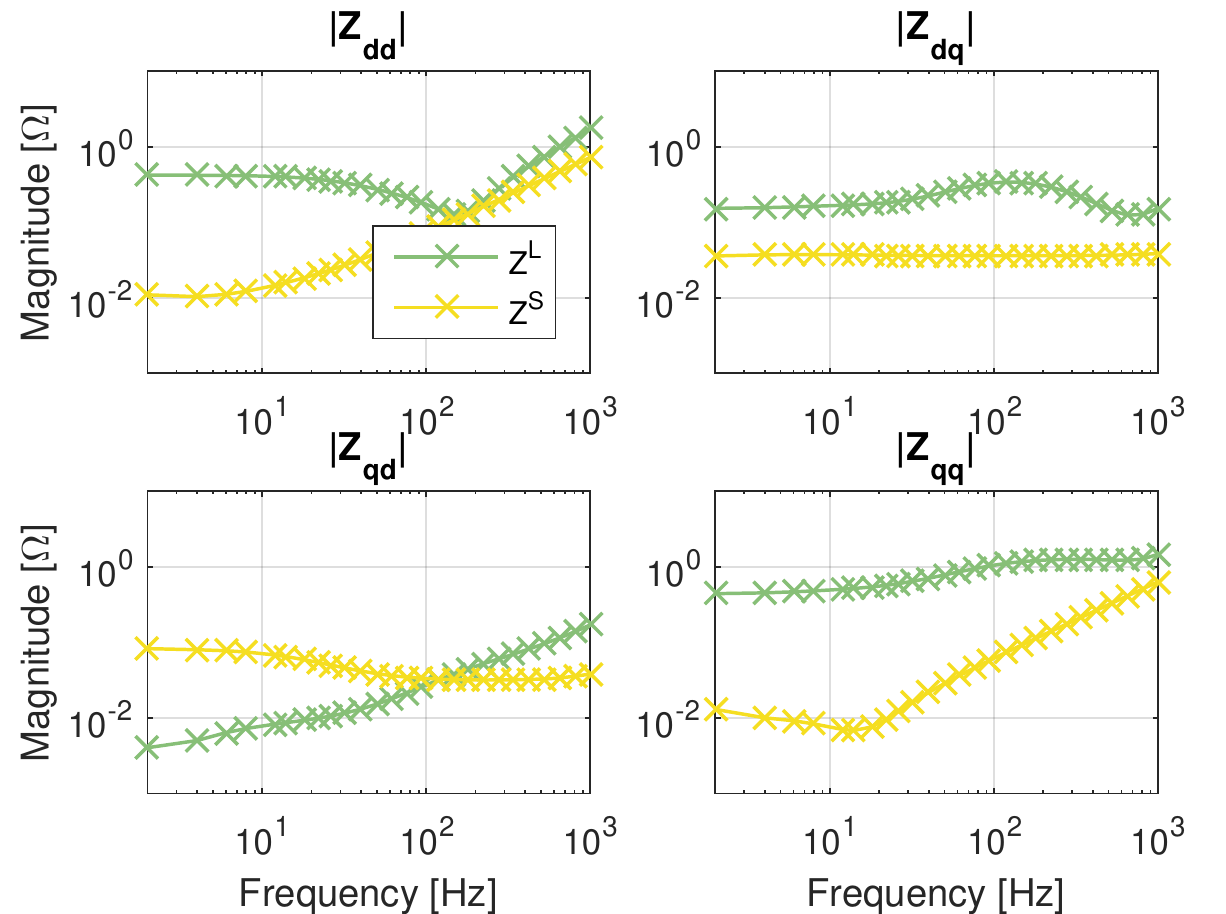}
     \caption{\textbf{Case A1} \textit{dq}-domain impedances for both subsystems}
     \label{fig:Result_Adq}
\end{figure}

\begin{figure}[ht!]
     \centering
     \includegraphics[width=0.48\textwidth]{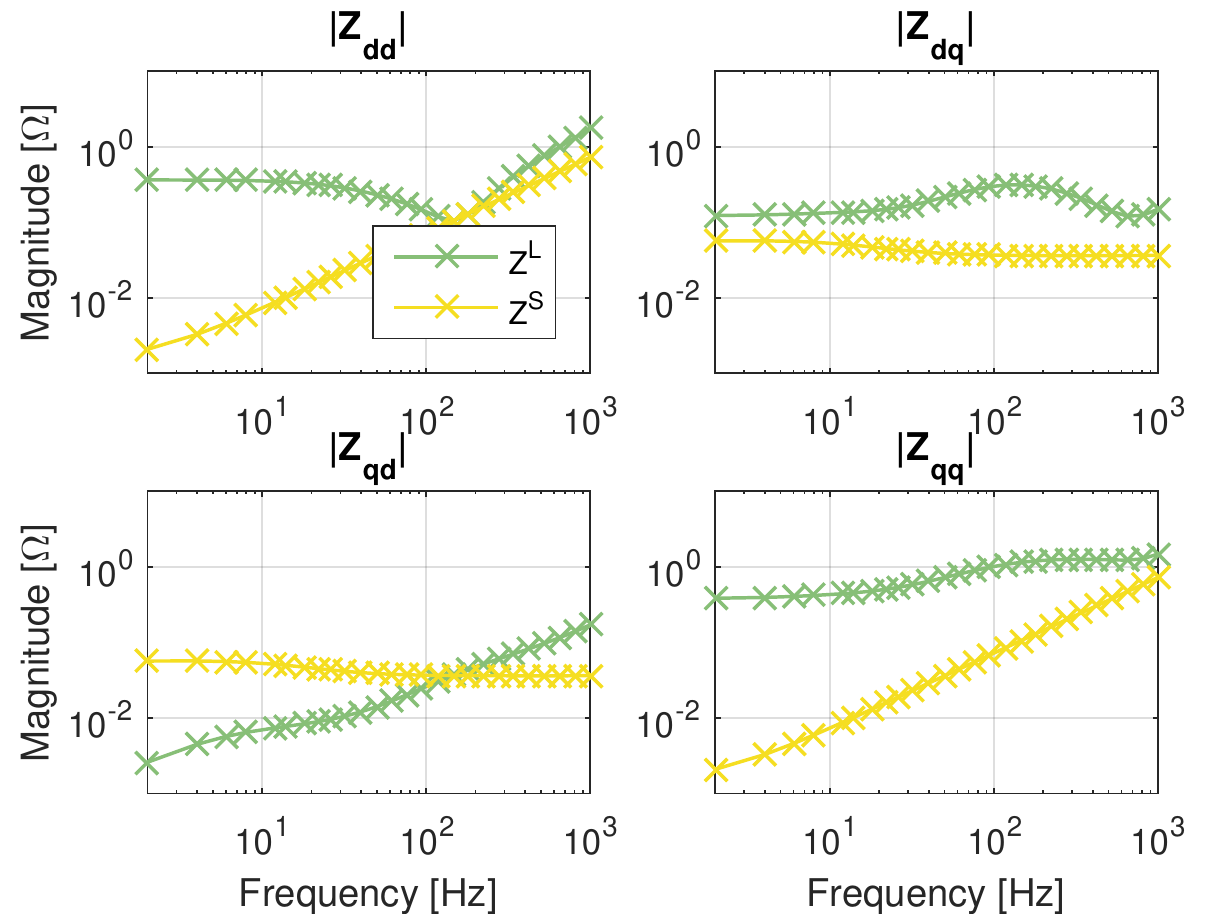}
     \caption{\textbf{Case A2} \textit{dq}-domain impedances for both subsystems}
     \label{fig:Result_Bdq}
\end{figure}

\begin{figure}[ht!]
     \centering
     \includegraphics[width=0.48\textwidth]{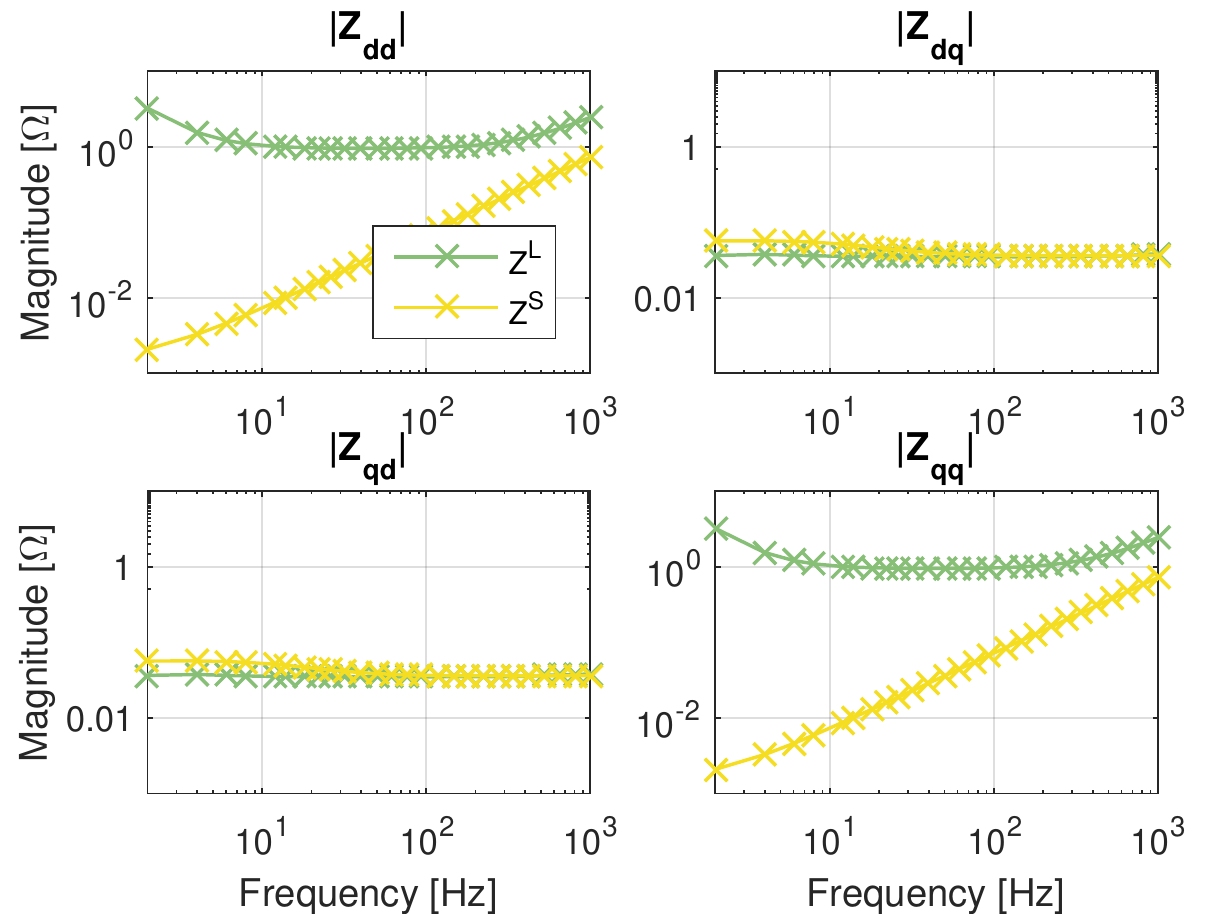}
     \caption{\textbf{Case B} \textit{dq}-domain impedances for both subsystems}
     \label{fig:Result_Cdq}
\end{figure}

\begin{figure}[ht!]
     \centering
     \includegraphics[width=0.48\textwidth]{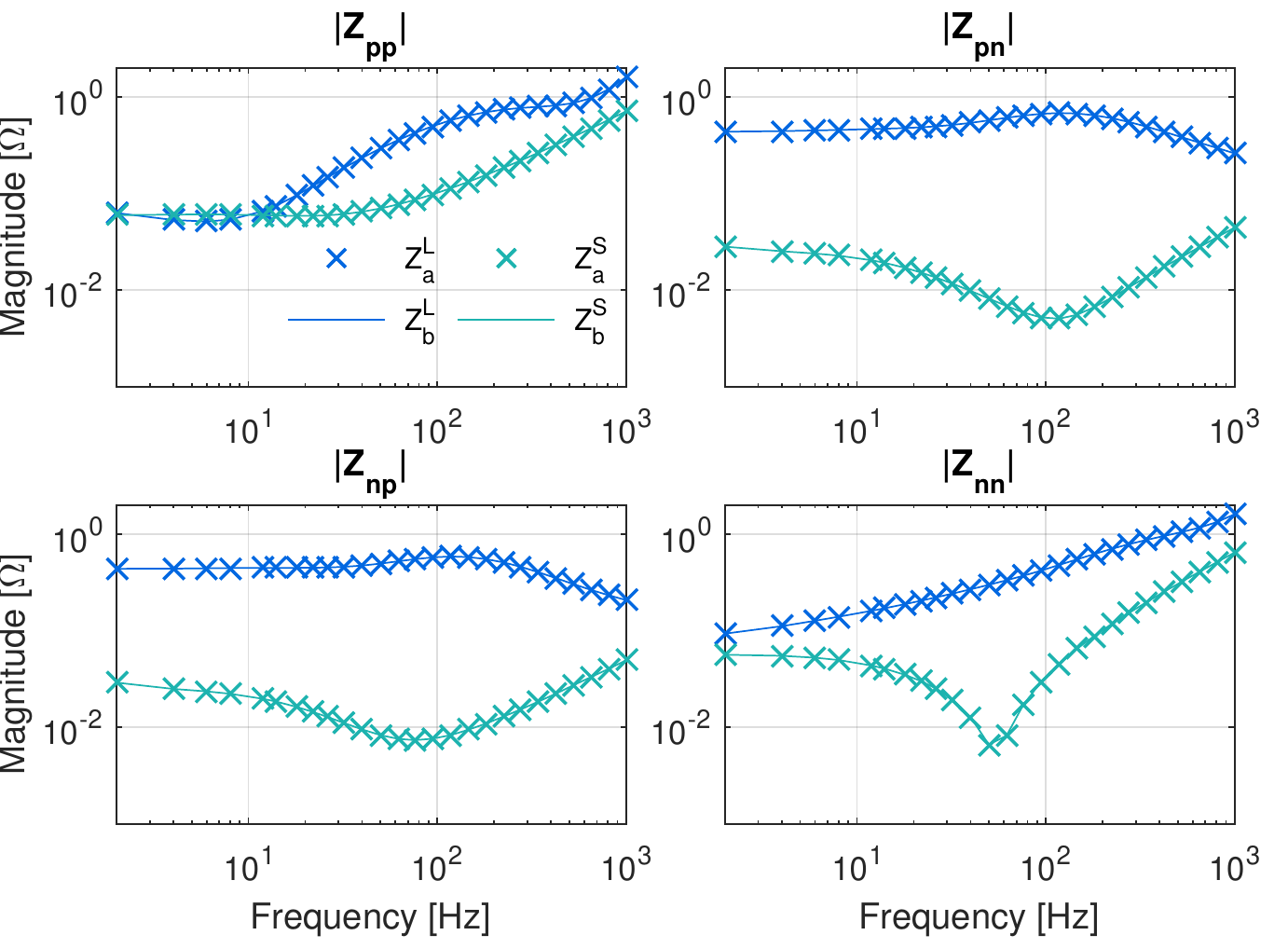}
     \caption{\textbf{Case A1} sequence domain impedances for both subsystems. \textit{a} and \textit{b} indicate two calculation methods.}
     \label{fig:Result_Apn}
\end{figure}

\begin{figure}[ht!]
     \centering
     \includegraphics[width=0.48\textwidth]{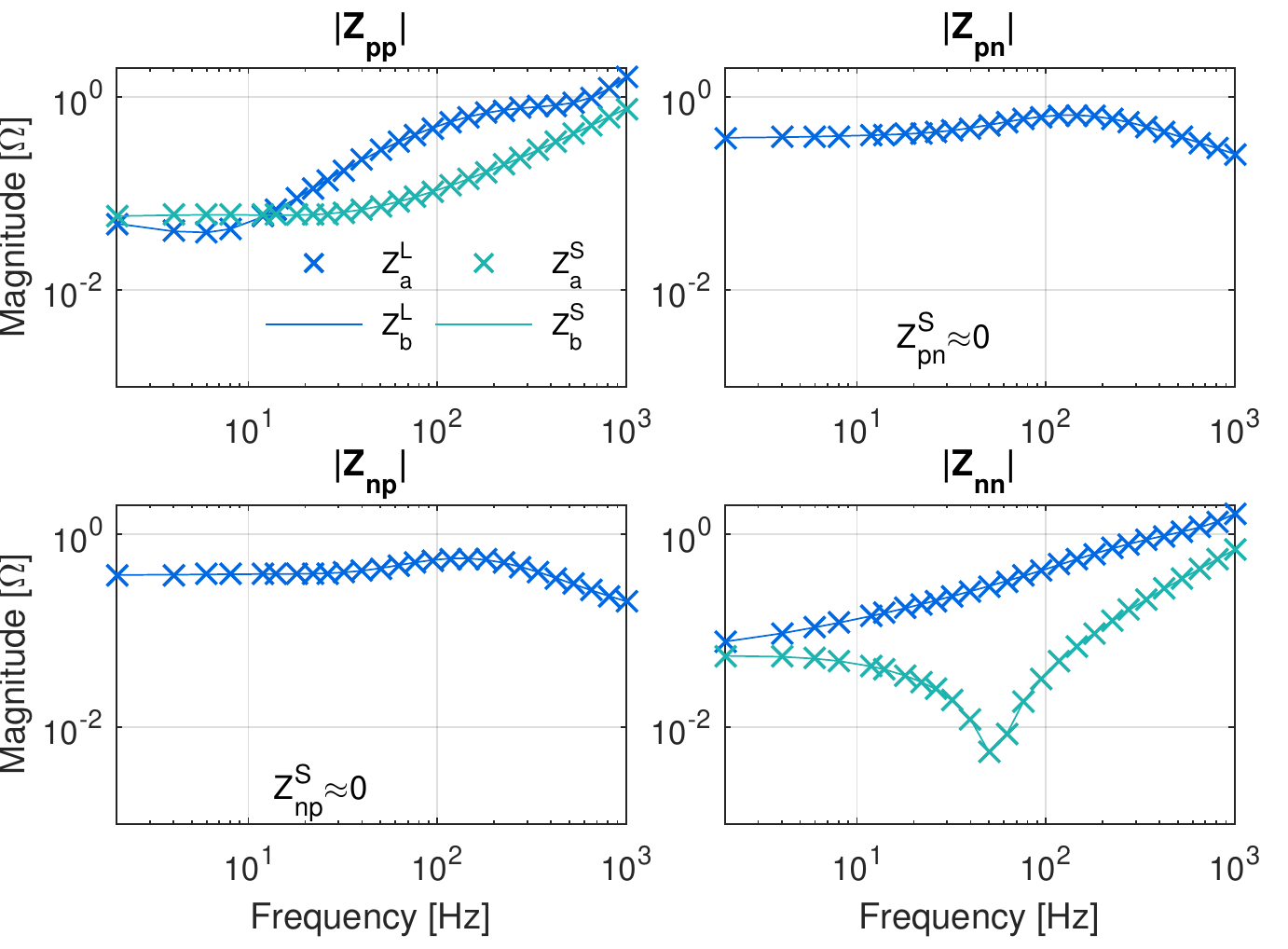}
     \caption{\textbf{Case A2} sequence domain impedances for both subsystems. \textit{a} and \textit{b} indicate two calculation methods.}
     \label{fig:Result_Bpn}
\end{figure}

\begin{figure}[ht!]
     \centering
     \includegraphics[width=0.48\textwidth]{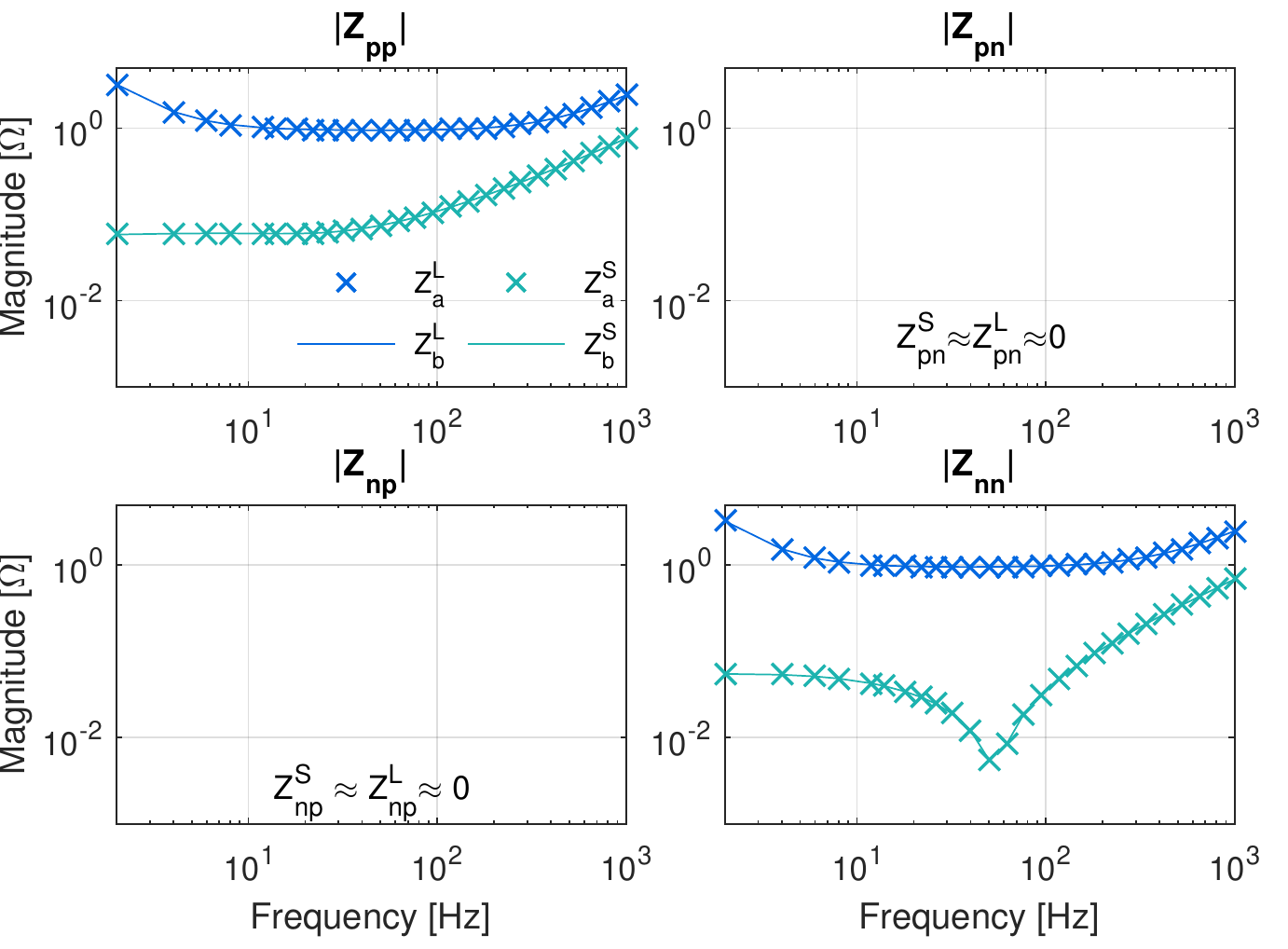}
     \caption{\textbf{Case B} sequence domain impedances for both subsystems. \textit{a} and \textit{b} indicate two calculation methods.}
     \label{fig:Result_Cpn}
\end{figure}

The resulting impedance curves for the three cases are shown in Figures \ref{fig:Result_Adq}-\ref{fig:Result_Cdq} for the \textit{dq}-domain, and in Figures \ref{fig:Result_Apn}-\ref{fig:Result_Cpn} for the modified sequence domain. Only magnitude is presented in these figures for simplicity, but is has been verified that the angles are consistent with the conclusions. Impedances for the load and source subsystem are plotted in the same graph. In the sequence domain plots, impedances have been obtained in two ways, denoted by subscript \textit{a} and \textit{b}:

\begin{itemize}
\item $Z_a$: Direct simulation of $\bm{Z_{pn}}$ using (\ref{eq:solve_Zpn})
\item $Z_b$: Based on simulated $\bm{Z_{dq}}$ from (\ref{eq:solve_Zdq}) and the transform given by by $\bm{Z_{pn}}=A_Z \cdot \bm{Z_{dq}} \cdot A_Z^{-1}$ (\ref{eq:Z_pn_2})
\end{itemize}

The following observations support the claims in the previous sections:

\begin{itemize}
\item The two ways of obtaining modified sequence domain impedances ($Z_a$ and $Z_b$) produce identical results in all cases. This confirms the transformation relationship from \textit{dq}- to sequence domain (\ref{eq:Z_pn_2}).
\item In \textbf{Case A1}, there are no symmetries in the impedance curves, neither in \textit{dq} nor sequence domain. This is expected since both subsystems have mirror frequency coupling.
\item In \textbf{Case A2}, $Z_{pn}^S \approx Z_{np}^S \approx 0$, as expected since the source subsystem is MFD (see also (\ref{eq:Z_mfd})).
\item In \textbf{Case B}, $Z_{pn} \approx Z_{np} \approx 0$ for both subsystems. This is expected since both subsystems are MFD. It can also be observed that $Z_{dq}=-Z_{qd}$ and $Z_{dd}=Z_{qq}$ for both subsystems, again according to (\ref{eq:Z_mfd}).
\end{itemize}

\clearpage
\newpage

\clearpage

\subsection{Simulation results - Modified vs. original}
\label{subsec:sim2}
\begin{figure}[ht!]
     \centering
     \includegraphics[width=0.45\textwidth]{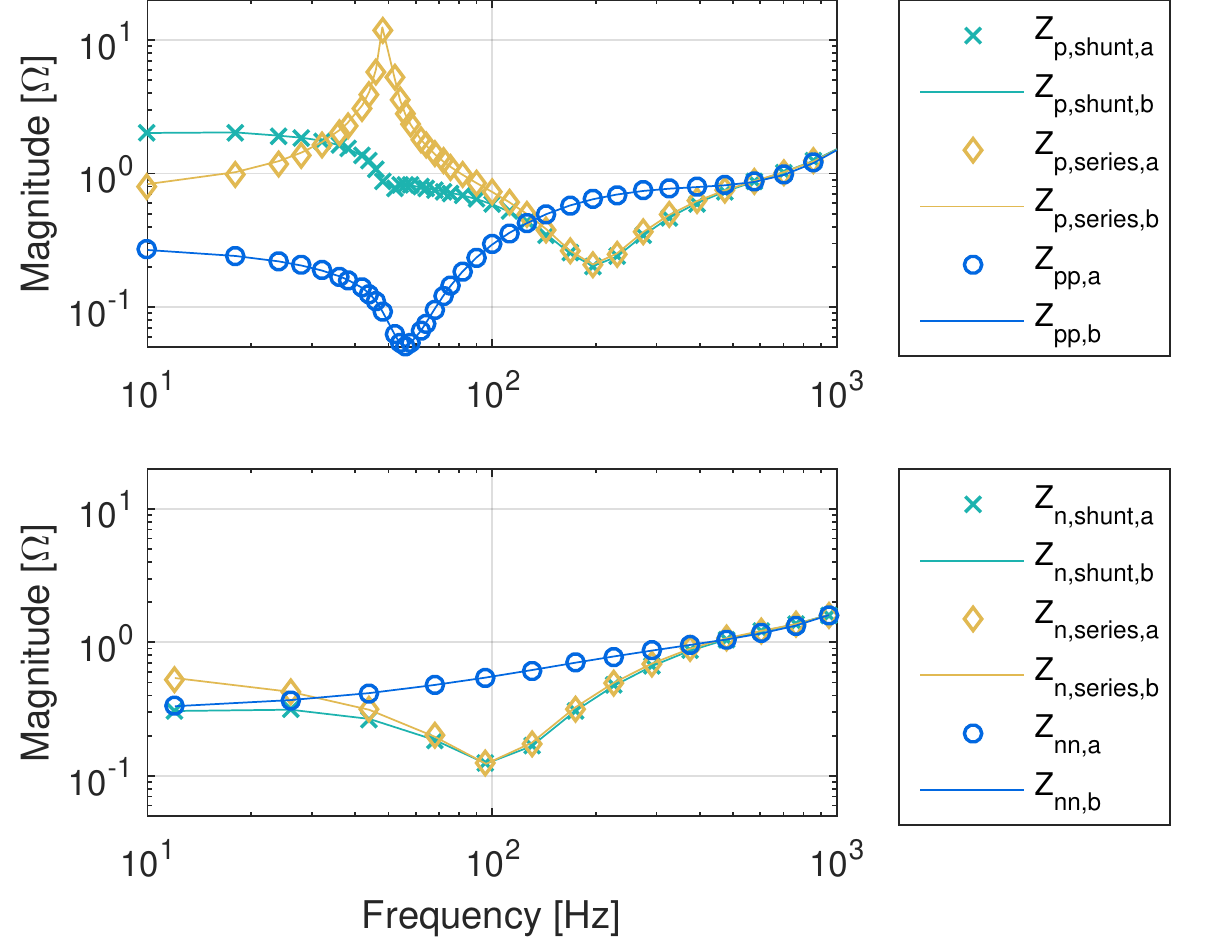}
     \caption{Comparison of \textit{load subsystem} original and modified sequence domain impedances in \textbf{Case A1}}
     \label{fig:Result_A_orig}
\end{figure}

\begin{figure}[ht!]
     \centering
     \includegraphics[width=0.45\textwidth]{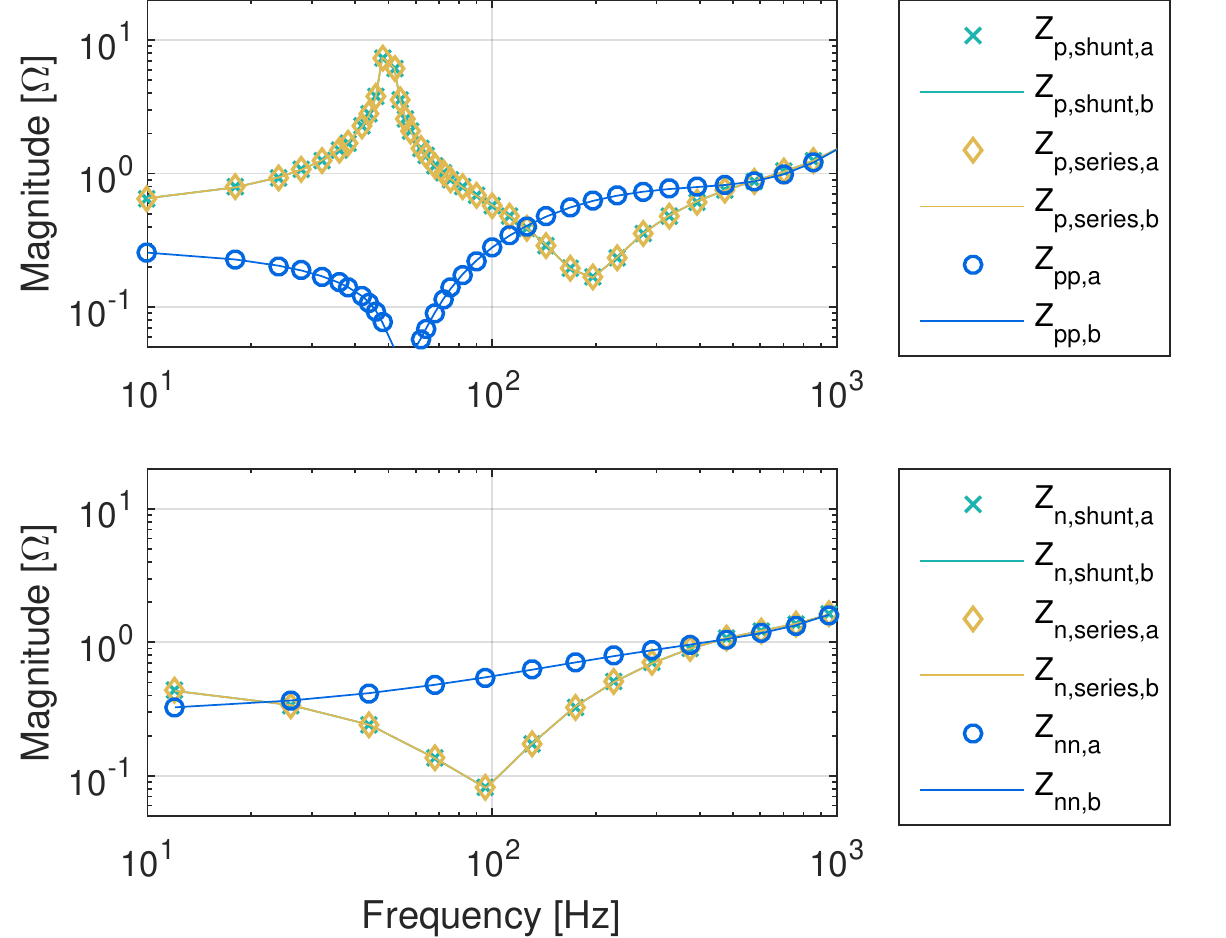}
     \caption{Comparison of \textit{load subsystem} original and modified sequence domain impedances in \textbf{Case A2}}
     \label{fig:Result_B_orig}
\end{figure}

\begin{figure}[ht!]
     \centering
     \includegraphics[width=0.45\textwidth]{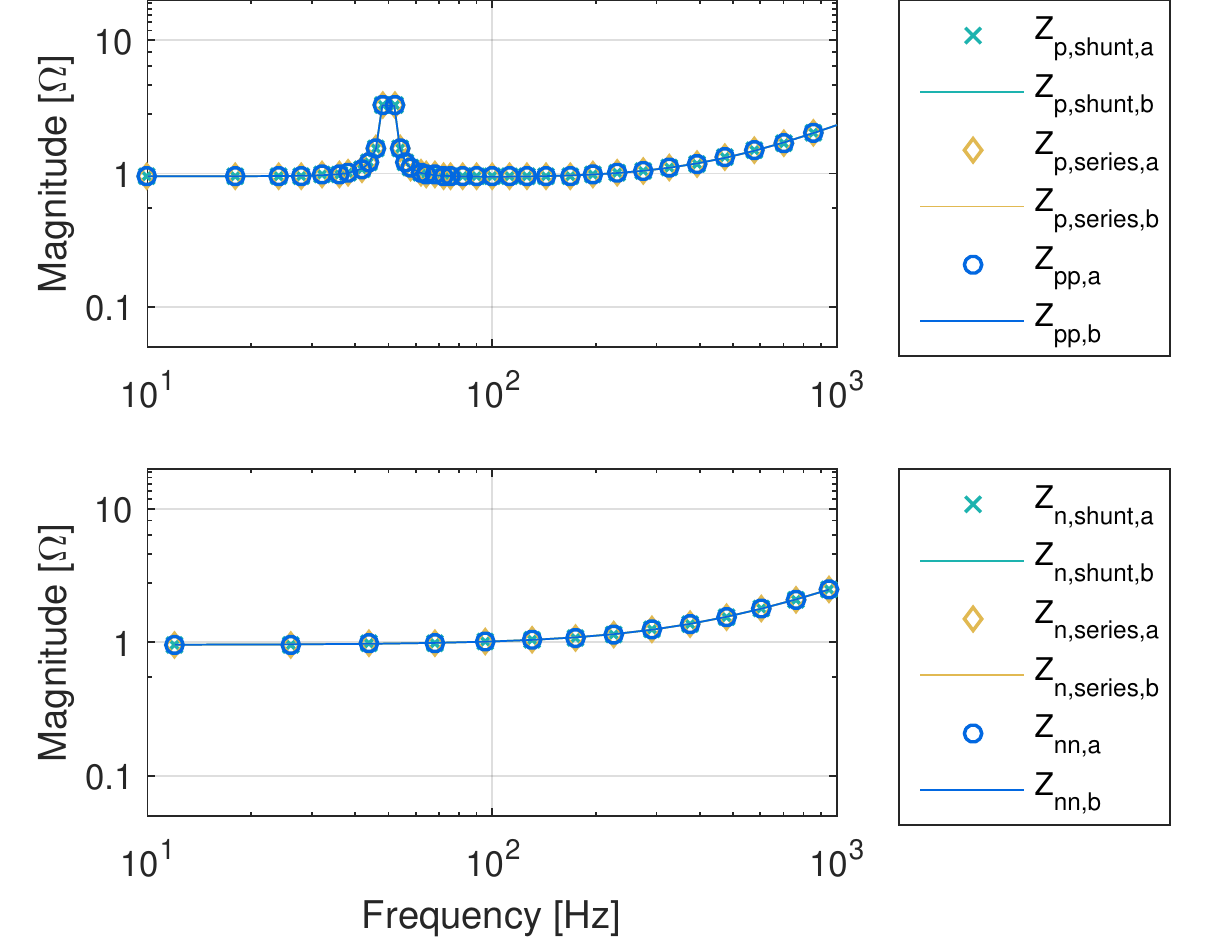}
     \caption{Comparison of \textit{load subsystem} original and modified sequence domain impedances in \textbf{Case B}}
     \label{fig:Result_C_orig}
\end{figure}

The purpose of this section is to investigate the validity of the equations derived in Appendix \ref{sec:origrelations}, and to determine how the original and modified sequence domain impedances relate to each other. Figure \ref{fig:Result_A_orig}-\ref{fig:Result_C_orig} presents the same comparison for each of the three simulation cases. Only the \textit{load subsystem} impedance is included in the comparison. The original sequence domain impedances, $Z_p$ and $Z_n$, are compared with the diagonal elements in the modified sequence domain impedance matrix, $Z_{pp}$ and $Z_{nn}$. Furthermore, the original sequence domain impedances are simulated both by shunt current and series voltage injection. Additionally, they are estimated in two ways to validate the equations derived in Appendix \ref{sec:origrelations}. Table \ref{tab:legend_exp} summarizes the notation and estimation methods.

\begin{table}[h]
\begin{center}
\caption{Explanation of the legends in Figures \ref{fig:Result_A_orig}-\ref{fig:Result_C_orig}}
\label{tab:legend_exp}
\begin{tabular}{ r | l | l}
Legend notation & Eq. notation & Estimation mehtod \\\hline
$Z_{p,shunt,a}$ & $Z_p^L\Big|_{shunt}$ & Direct simulation using (\ref{eq:Zpn_orig}) \\
$Z_{p,shunt,b}$ & $Z_p^L\Big|_{shunt}$ &
\begin{tabular}[c]{@{}c@{}}Calculation based on simulated $\bm{Z_{pn}}$\\from (\ref{eq:solve_Zpn}) and the formula (\ref{eq:Zp_shunt})\end{tabular}
  \\
$Z_{p,series,a}$ & $Z_p^L\Big|_{series}$ & Direct simulation using (\ref{eq:Zpn_orig}) \\
$Z_{p,series,b}$ & $Z_p^L\Big|_{series}$ & 
\begin{tabular}[c]{@{}c@{}}Calculation based on simulated $\bm{Z_{pn}}$\\from (\ref{eq:solve_Zpn}) and the formula (\ref{eq:Zp_series})\end{tabular}
\\
$Z_{pp,a}$ &  $Z_{pp}^L$ & Direct simulation using (\ref{eq:solve_Zpn})\\
$Z_{pp,b}$ & $Z_{pp}^L$ & 
\begin{tabular}[c]{@{}c@{}}Calculation based simulated $\bm{Z_{dq}}$
\\from (\ref{eq:solve_Zdq}) and the formula (\ref{eq:Z_pn_2})\end{tabular}
\end{tabular}
\end{center}
\end{table}

First, it can be noted that the methods based on direct simulations always overlap with the calculated one for all three cases, which validates (\ref{eq:Zp_shunt}), (\ref{eq:Zp_series}) and (\ref{eq:Z_pn_2}). 

In \textbf{Case A1}, the original impedances obtained by shunt and series injection are not equal. This is expected based on the difference between (\ref{eq:Zp_shunt}) and (\ref{eq:Zp_series}). The difference is noticeable at frequencies below \textit{100 Hz}. It is also observed that the modified sequence impedances $Z_{pp}$ and $Z_{nn}$ deviate substantially from the original ones at frequencies up to $\approx 500 Hz$; however, after this point they are close to equal. Consequently, the load subsystem can be assumed to be MFD for frequencies above \textit{500 Hz}.

In \textbf{Case A2}, it is clear that the load subsystem impedances obtained from shunt and series injection are equal. This is consistent with (\ref{eq:Zp_oneMFD}), because the source subsystem is MFD in this case. The same equation give $Z_{pp}\neq Z_p$ and $Z_{nn}\neq Z_n$, which is verified by the figure. However, the difference is close to zero at frequencies above $\approx 500 Hz$, similar to \textbf{Case A1}.

In \textbf{Case B}, all impedance estimates coincide. From (\ref{eq:bothMFD}) we have that $Z_p=Z_{pp}$ and $Z_n=Z_{nn}$ whenever $Z_{pn}=Z_{np}=0$. Both systems are now MFD, and hence the assumption of decoupled sequence domain is valid.

\newpage

\subsection{Simulation results - Generalized Nyquist Criterion}
\begin{figure}[ht!]
     \centering
     \includegraphics[width=0.35\textwidth]{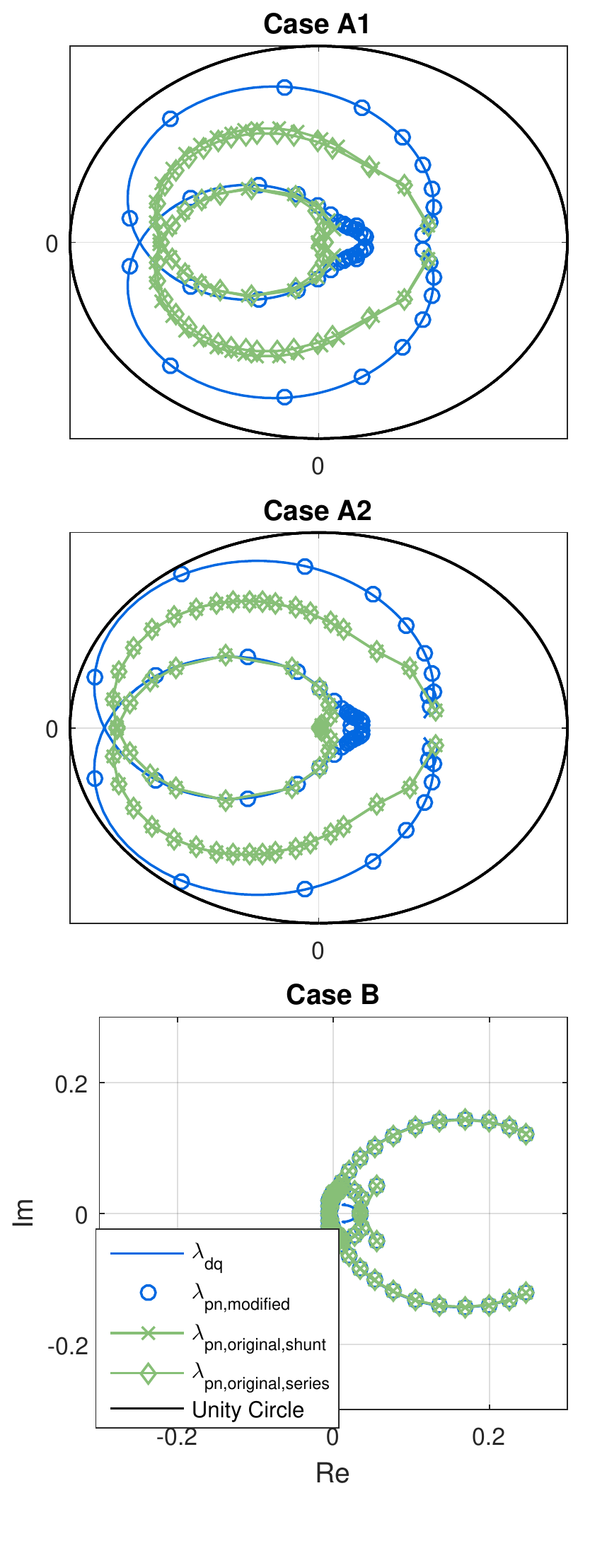}
     \caption{Comparison of Nyquist plots for cases A1, A2 and B}
     \label{fig:Result_GNC}
\end{figure}

It has been shown in section \ref{subsec:GNC} that the Generalized Nyquist Criterion will give same result in the \textit{dq}-domain and the modified sequence domain. Furthermore, it has been shown that the stability analysis based on original sequence domain impedances will give different results unless both subsystems are MFD. This has been investigated by applying the GNC to the impedance curves found in Figure \ref{fig:Result_Adq} to Figure \ref{fig:Result_C_orig}. The resulting Nyquist plots are presented in Figure \ref{fig:Result_GNC}. Note that only the most critical eigenvalue is plotted, corresponding to the operating point of $i_{sd}=1.1 pu$.

In all cases, the \textit{dq}-domain gives exactly the same result as the modified sequence domain. On the other hand, the original sequence domain impedances do not give the same Nyquist plot in Case A1 and A2. In Case B all methods give the same Nyquist curves.

\subsection{Time-domain analysis}
\begin{figure}[ht!]
     \centering
     \includegraphics[width=0.49\textwidth]{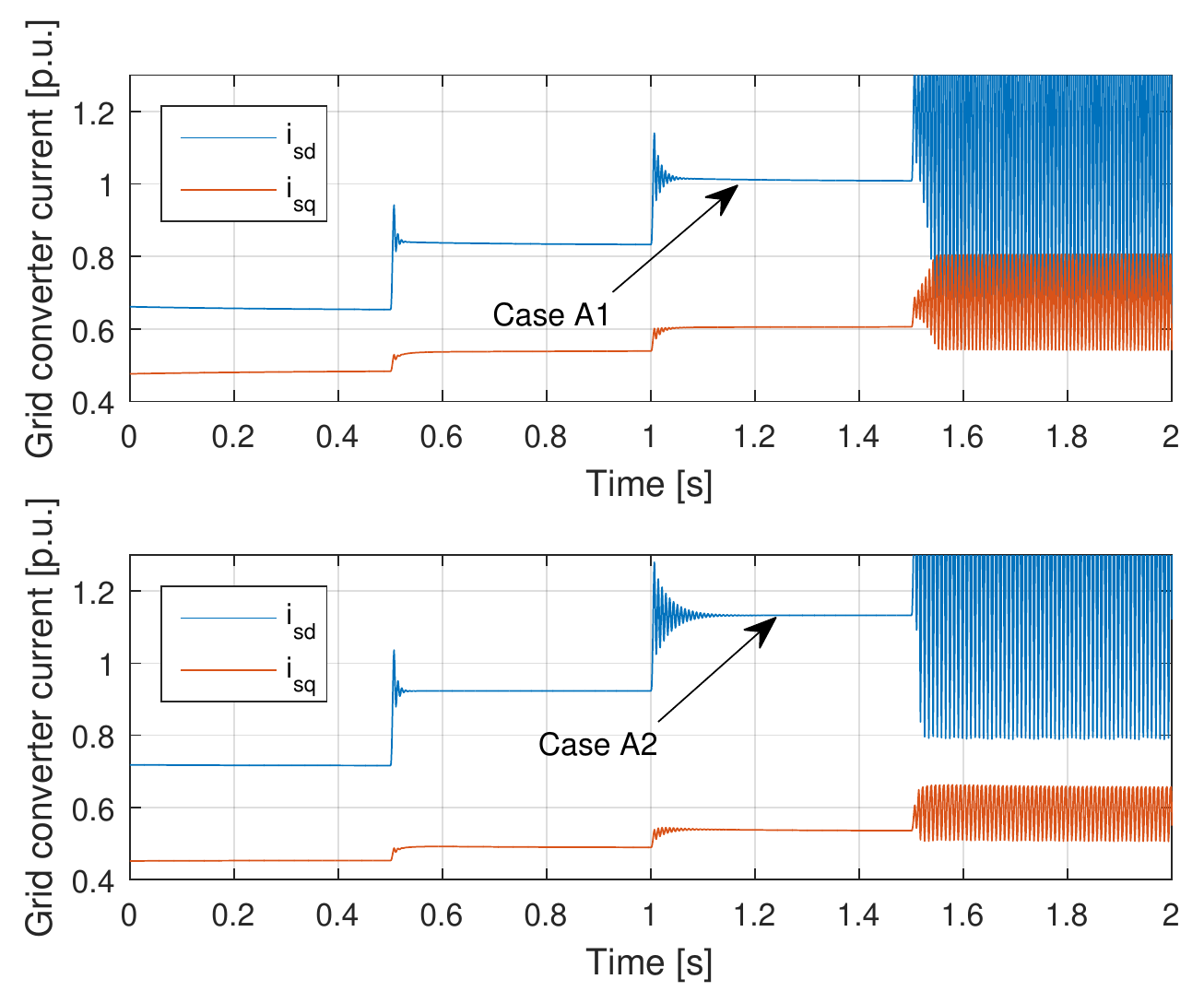}
     \caption{Time-domain analysis of Case A1 and Case A2. Grid converter \textit{dq}-currents during stepwise increase in $I_{dc}$. Arrows indicate the operation point used for impedance calculations.}
     \label{fig:Result_timedomain}
\end{figure}

To complement the stability analysis from the previous section, a time-domain simulation has been conducted and presented in Figure \ref{fig:Result_timedomain}. Cases A1 and A2 are used for the simulation, see Figure \ref{fig:CaseA} for the system block diagram. The power consumed by the load converter is stepwise increased by applying step changes to $I_{dc}$. Figure \ref{fig:Result_timedomain} shows the resulting \textit{dq}-currents at the \textit{source converter}. The transient oscillations in both $i_{sd}$ and $i_{sq}$ are gradually increasing until instability occurs at the reference value of $1.2$ $p.u.$. Note that a reference value of $1.1$ $p.u.$ is the basis for the Nyquist plot in Figure \ref{fig:Result_GNC} as indicated by arrows. This operation point has poorly damped oscillations, supporting the fact that the Nyquist plot is close to encircling the point $(-1,0)$. It can be observed that Case A2 has stronger oscillations, which is also indicated by its Nyquist plot being closer to $(-1,0)$ compared to Case A1.

\section{Discussion}
After the derivations and proofs presented in the previous sections it is useful to discuss the advantages and disadvantages of the two impedance domains. In this discussion, the modified sequence domain is assumed because the original definition has been shown to be ambiguous and can be inaccurate for systems that are not MFD.

The general statement is that \textit{the two domains are equivalent}. The impedance matrix in one domain can be obtained from the other through a linear transformation where the eigenvalues are preserved. Hence, both domains should provide the same results. However, from a more practical viewpoint, there are a few advantages with the sequence domain over \textit{dq}, especially when impedances are obtained from simulation or measurements:
\begin{enumerate}
\item There is no need to perform \textit{dq}-transformations to any measured signal, and hence no need for a reference transformation angle.
\item The off-diagonal terms in the sequence domain impedance matrix will often have low values, and are equal to zero for a MFD system. In other words, the sequence domain impedance matrix is “close to decoupled”.
\item It can be argued that the sequence domain can be intuitively associated to a physical meaning  and is less abstract than the \textit{dq}-domain.
\end{enumerate}

The main advantage with the \textit{dq}-domain is that most previous studies on the control and stability of power electronic converters have been performed in \textit{dq}-coordinates. Regarding analytical impedance models, it is also possible that the derivations are less complex in the \textit{dq}-domain, but this has not been yet extensively investigated. 

In most of the previous work on stability analysis based on the Nyquist criterion, it has been argued that off-diagonal impedance matrix elements can be neglected. Based on the impedance matrix structure for MFD systems (\ref{eq:Z_mfd}), it is clear that such simplifications are correct in the sequence domain but not in the \textit{dq}-domain.

The paper has shown that original sequence domain impedances depends on injection type (shunt vs. series) when \textit{both} subsystems are non-MFD. In general, the load subsystem prefer series injection, while the source subsystem prefer shunt injection. This is due to the fact that the load subsystem has higher impedance at most frequencies. When choosing injection type for a given system, the system with more mirror frequency coupling should be prioritized. It is expected that this is often the load subsystem, in this case series injection is more accurate.


\section{Conclusion}
Stability analysis of AC power electronics systems through frequency dependent impedance equivalents is a relatively new field of research. Both \textit{dq-domain} and \textit{sequence domain} analysis have been reported in previous works. However, limited effort has been dedicated to the understanding of their equivalence with respect to the stability estimates they provide. This paper attempts to contribute in this direction by reporting the following findings:
\begin{enumerate}
\item A modified definition for the sequence domain impedance matrix, which extends the original sequence domain impedance definition. The extension is related with the ability to account for induced frequency components shifted by twice the fundamental frequency

\item The relationship between the well-established \textit{dq}-domain impedance matrix and the modified sequence domain impedance matrix. This can be viewed as a linear transformation where many essential properties are preserved.

\item The choice of impedance domain does not affect the analysis of stability using the Generalized Nyquist Criterion (GNC).

\item Definition of the terms \textit{mirror frequency effect} and \textit{Mirror Frequency Decoupled} (MFD) systems.

\item The modified sequence domain impedance matrix is diagonal for MFD systems. The \textit{dq}-domain impedance matrix is skew symmetric.

\item The \textit{dq}-domain impedance matrix can be obtained from single measurements without the need for matrix inversions when both subsystems are MFD. 

\item The original sequence domain impedance is shown to be ambiguous in the general case; it depends on injection type, and the source state variables appear in the load impedance expression, and vice versa.

\item The relationship between the original and the modified sequence domain impedance was derived under the assumption of both ideal series and ideal shunt injection.

\item The original sequence domain impedance no longer depends on injection type when \textit{one} of the subsystems is MFD. However, the source state variables still appear in the load impedance expression (assuming the source is MFD).

\end{enumerate}

All equations were derived mathematically on a general basis, and were verified by numeric simulations. Points 1-5 were examined in simulation section \ref{subsec:sim1}, whereas points 6-9 were examined in simulation section \ref{subsec:sim2}.

\newpage


\small
\bibliographystyle{IEEEtran}
\bibliography{PhD-arbeid}


\appendix

\subsection{Parameter values used in simulations}
\label{app:params}

\begin{table}[h!]
    \centering
    \begin{tabular}{l | l | l}
         $V_{base}$=  690 V   &   $S_{base}$ = $1$ $MW$  & $V_{dc,base}$=  $1400$ $V$ \\
        $V_{Sdc}$ = 1 $p.u.$ & $V_{Ldc}^*$ = 1 $p.u.$   &   $Z_{S}$ = $0.007+j0.15$ $p.u.$ \\
        $C_{dc}=11.5$ $mF$ & $I_{dc}=1.1$ $p.u.$ & $Z_{L}$ = $0.02+j0.25$ $p.u.$ \\
        $L_{vd}$=  $0.0$ $p.u.$  &   $K_{pvd}$ = $1$ $p.u.$  & $T_{ivd}$=  $0.1$ $s$ \\
        $L_{vq}$=  $0.2$ $p.u.$  &   $K_{pvq}$ = $1.3$ $p.u.$  & $T_{ivq}$=  $0.2$ $s$ \\
        $K_{pid}$ = $1.59$ $p.u.$  & $T_{iid}$=  $0.047$ $s$ & $f_{n}$=  $50$ $Hz$ \\
        $K_{piq}$ = $2.07$ $p.u.$  & $T_{iiq}$=  $0.033$ $s$ & $K_{pvdc}$ = $8.33$ $p.u.$ \\
        $T_{ivdc}$ = $0.0036$ $s$  &  $v_d^*=1.0$ $p.u.$ & $v_q^*=0.0$ $p.u.$ \\
        $i_{Lq}^*=0.4$ $p.u.$ & & \\
    \end{tabular}
    \caption{Parameter values applied in \textbf{Case A1}}
    \label{tab:param_A1}
\end{table}

\begin{table}[h!]
    \centering
    \begin{tabular}{l | l | l}
         $V_{base}$=  690 V   &   $S_{base}$ = $1$ $MW$  & $V_{dc,base}$=  $1400$ $V$ \\
        $V_{Sdc}$ = 1 $p.u.$ & $V_{Ldc}^*$ = 1 $p.u.$   &   $Z_{S}$ = $0.007+j0.15$ $p.u.$ \\
        $C_{dc}=11.5$ $mF$ & $I_{dc}=1.1$ $p.u.$ & $Z_{L}$ = $0.02+j0.25$ $p.u.$ \\
        \textcolor{red}{$L_{vd}$=  $0.1$ $p.u.$}  &   $K_{pvd}$ = $1$ $p.u.$  & $T_{ivd}$=  $0.1$ $s$ \\
        \textcolor{red}{$L_{vq}$=  $0.1$ $p.u.$}  &   \textcolor{red}{$K_{pvq}$ = $1$ $p.u.$}  & \textcolor{red}{$T_{ivq}$=  $0.1$ $s$ }\\
        $K_{pid}$ = $1.59$ $p.u.$  & $T_{iid}$=  $0.047$ $s$ & $f_{n}$=  $50$ $Hz$ \\
        $K_{piq}$ = $2.07$ $p.u.$  & $T_{iiq}$=  $0.033$ $s$ & $K_{pvdc}$ = $8.33$ $p.u.$ \\
        $T_{ivdc}$ = $0.0036$ $s$  &  $v_d^*=1.0$ $p.u.$ & $v_q^*=0.0$ $p.u.$ \\
        $i_{Lq}^*=0.4$ $p.u.$ & & \\
    \end{tabular}
    \caption{Parameter values applied in \textbf{Case A2}. Parameters in \textcolor{red}{RED} are different from Case A1.}
    \label{tab:param_A2}
\end{table}

\begin{table}[h!]
    \centering
    \begin{tabular}{l | l | l}
         $V_{base}$=  690 V   &   $S_{base}$ = $1$ $MW$  & $V_{dc,base}$=  $1400$ $V$ \\
        $V_{Sdc}$ = 1 $p.u.$ & $V_{Ldc}$ = 1 $p.u.$   &   $Z_{S}$ = $0.007+j0.15$ $p.u.$ \\
        $C_{dc}=11.5$ $mF$ & \textcolor{red}{$i_{Ld}^*=1.1$ $p.u.$} & $Z_{L}$ = $0.02+j0.25$ $p.u.$ \\
        $L_{vd}$=  $0.1$ $p.u.$  &   $K_{pvd}$ = $1$ $p.u.$  & $T_{ivd}$=  $0.1$ $s$ \\
        $L_{vq}$=  $0.1$ $p.u.$  &   $K_{pvq}$ = $1$ $p.u.$  & $T_{ivq}$=  $0.1$ $s$ \\
        $K_{pid}$ = $1.59$ $p.u.$  & $T_{iid}$=  $0.047$ $s$ & $f_{n}$=  $50$ $Hz$ \\
        \textcolor{red}{$K_{piq}$ = $1.59$ $p.u.$}  & \textcolor{red}{$T_{iiq}$=  $0.047$ $s$} & $i_{Lq}^*=0.4$ $p.u.$\\
        $v_d^*=1.0$ $p.u.$ & $v_q^*=0.0$ $p.u.$ & \\
    \end{tabular}
    \caption{Parameter values applied in \textbf{Case B}. Parameters in \textcolor{red}{RED} are different from Case A2.}
    \label{tab:param_B}
\end{table}

\subsection{Proof of equal determinants}\label{app:det}
The following relation prove that the determinant of $\bm{Z_{dq}}$ is always equal to the determinant of $\bm{Z_{pn}}$:
\begin{align}
\text{det} (\bm{Z_{pn}})
&= \text{det}\left( A_Z \cdot \bm{Z_{dq}} \cdot A_Z^{-1} \right)
\nonumber \\
&=\text{det}(A_Z) \cdot \text{det}\left(\bm{Z_{dq}}\right) \cdot \text{det} (A_Z^{-1})
\nonumber \\
&=-j \cdot \text{det}\left(\bm{Z_{dq}}\right) \cdot j 
= \text{det}\left(\bm{Z_{dq}}\right)
\end{align}

\subsection{Proof of MFD impedance matrices relations (\ref{eq:Z_mfd})}\label{app:mfd}
When a subsystem is assumed MFD, the off-diagonal elements in $\bm{Z_{pn}}$ are equal to zero by definition. Hence $Z_{pn}=Z_{np}=0$. Substituting from (\ref{eq:Z_pn}) then gives:
\begin{align}
Z_{pn}&=\frac{1}{2}
\begin{bmatrix}1 & -j\end{bmatrix}\bm{Z_{dq}}\begin{bmatrix}1 \\ -j \end{bmatrix}=0
\nonumber \\
Z_{np}&=\frac{1}{2}
\begin{bmatrix}1 & j\end{bmatrix}\bm{Z_{dq}}\begin{bmatrix}1 \\ j \end{bmatrix}=0
\end{align}

Expanding these expressions by substituting from (\ref{eq:Ohms_dq}) gives:
\begin{align}
Z_{dd}&=Z_{qq}=Z_x \nonumber \\
Z_{dq}&=-Z_{qd}=Z_y \nonumber \\
Z_{pp} &= \frac{Z_x-jZ_y}{2} \nonumber \\
Z_{nn} &= \frac{Z_x + jZ_y}{2}
\end{align}

$Z_x$ and $Z_y$ are defined in (\ref{eq:Z_mfd}).

\subsection{Relationship between modified and original sequence domain impedance definitions}
\label{sec:origrelations}

\subsubsection{General case}
\label{subsec:general}

The relationship between the modified sequence domain impedance definition (\ref{eq:Z_pn_2}) and the original (\ref{eq:Zpn_orig}) can be derived by solving (\ref{eq:Ohms_pn}) for the source and load subsystem simultaneously. When specifying the set of equations, one must choose between:
\begin{itemize}
\item Shunt current or series voltage injection
\end{itemize}
.. and between
\begin{itemize}
\item Positive or negative sequence injection
\end{itemize}

One should choose positive sequence injection in order to find the positive sequence impedance $Z_p$, and negative sequence injection to find $Z_n$. The following set of equations should be solved to obtain the impedance $Z_p$ for shunt current injection. This is equivalent to solving the circuit presented in Figure \ref{fig:MFE_illustration}.

\begin{align}
V_p^L &= I_p^L Z_{pp}^L + I_n^L Z_{pn}^L \nonumber \\
V_n^L &= I_p^L Z_{np}^L + I_n^L Z_{nn}^L \nonumber \\
V_p^S &= I_p^S Z_{pp}^S + I_n^S Z_{pn}^S \nonumber \\
V_n^S &= I_p^S Z_{np}^S + I_n^S Z_{nn}^S \nonumber \\
V_p^L &= V_p^S = V_p \nonumber \\
V_n^L &= V_n^S = V_n \nonumber \\
I_n^S &= -I_n^L \nonumber \\
Z_p^L\Big|_{shunt} &= \frac{V_p}{I_p^L} \nonumber \\
Z_p^S\Big|_{shunt} &= \frac{V_p}{I_p^S}
\label{eq:Zp_shunt_eqs}
\end{align}
The superscript \textit{L} denotes load subsystem, whereas \textit{S} denotes source subsystem. The first four equations are the Generalized Ohms Law with the modified sequence domain definition. The last five equations depend on the choice of injection type, as well as the choice of positive or negative sequence injection. The voltages in the two subsystems are equal if the injection is shunt type. Furthermore, the sum of negative sequence current must be zero because the injected perturbation is assumed to be pure positive sequence. Solving (\ref{eq:Zp_shunt_eqs}) gives the following \textit{original} impedance:

\begin{align}
Z_p^L\Big|_{shunt} &= \frac{V_p}{I_p^L} =
\frac{Z_{pp}^L D^S + Z_{pp}^S D^L}{D^S + Z_{nn}^L Z_{pp}^S - Z_{pn}^L Z_{np}^S}
\nonumber \\
Z_p^S\Big|_{shunt} &= \frac{V_p}{I_p^S} =
\frac{Z_{pp}^S D^L + Z_{pp}^L D^S}{D^L + Z_{nn}^S Z_{pp}^L - Z_{pn}^S Z_{np}^L}
\label{eq:Zp_shunt}
\end{align}

where $D^S$ and $D^L$ are the determinants of the source and load \textit{dq}-domain impedance matrices $\bm{Z_{dq}^S}$ and $\bm{Z_{dq}^L}$, respectively:
\begin{align}
D^S &= Z_{dd}^S Z_{qq}^S - Z_{dq}^S Z_{qd}^S \nonumber \\
D^L &= Z_{dd}^L Z_{qq}^L- Z_{dq}^L Z_{qd}^L
\end{align}
It is shown in Appendix \ref{app:det} that the determinant of $\bm{Z_{pn}}$ equals the determinant of $\bm{Z_{dq}}$.

A corresponding expression can be derived for series voltage positive sequence injection. The last five equations in (\ref{eq:Zp_shunt_eqs}) are then modified to:

\begin{align}
I_p^L &= -I_p^S = I_p \nonumber \\
I_n^L &= -I_n^S = I_n \nonumber \\
V_n^S &= V_n^L \nonumber \\
Z_p^L\Big|_{series} &= \frac{V_p^L}{I_p} \nonumber \\
Z_p^S\Big|_{series} &= \frac{V_p^S}{I_p} 
\label{eq:Zp_series_eqs}
\end{align}

Solving the set of equations gives:
\begin{align}
Z_p^L\Big|_{series} &= \frac{V_p^L}{I_p} =
\frac{Z_{pp}^L Z_{nn}^S - Z_{pn}^L Z_{np}^S + D^L}{Z_{nn}^S + Z_{nn}^L}
\nonumber \\
Z_p^S\Big|_{series} &= \frac{V_p^S}{I_p} =
\frac{Z_{pp}^S Z_{nn}^L - Z_{pn}^S Z_{np}^L + D^S}{Z_{nn}^L + Z_{nn}^S}
\label{eq:Zp_series}
\end{align}

Given that (\ref{eq:Zp_series}) clearly differs from (\ref{eq:Zp_shunt}), it can be concluded that the original sequence domain impedance is not well defined in the general case because it depends on the injection type. This has been illustrated by simulations in Figure \ref{fig:Result_A_orig}.

The additional equations for negative sequence are given in (\ref{eq:additional_Z}).

\begin{align}
Z_n^L\Big|_{shunt} &= \frac{V_n}{I_n^L} =
\frac{Z_{nn}^L D^S + Z_{nn}^S D^L}{D^S + Z_{pp}^L Z_{nn}^S - Z_{np}^L Z_{pn}^S}
\nonumber \\
Z_n^L\Big|_{series} &= \frac{V_n^L}{I_n} =
\frac{Z_{nn}^L Z_{pp}^S - Z_{np}^L Z_{pn}^S + D^L}{Z_{pp}^S + Z_{pp}^L}
\nonumber \\
Z_n^S\Big|_{shunt} &= \frac{V_n}{I_n^S} =
\frac{Z_{nn}^S D^L + Z_{nn}^L D^S}{D^L + Z_{pp}^S Z_{nn}^L - Z_{np}^S Z_{pn}^L}
\nonumber \\
Z_n^S\Big|_{series} &= \frac{V_n^S}{I_n} =
\frac{Z_{nn}^S Z_{pp}^L - Z_{np}^S Z_{pn}^L + D^S}{Z_{pp}^L + Z_{pp}^S}
\label{eq:additional_Z}
\end{align}

In section \ref{sec:sim} these analytic expressions are validated through a comparison where the original sequence impedances are obtained directly by simulation.

\subsubsection{Special case with one MFD subsystem}
In the special case where one subsystem is MFD, the expressions from the previous sections can be simplified. If the source subsystem is MFD, i.e. $Z_{pn}^S=Z_{np}^S=0$, then, (\ref{eq:Zp_shunt}), (\ref{eq:Zp_series}) and (\ref{eq:additional_Z}) are reduced to:
\begin{align}
Z_p^L\Big|_{shunt} &= Z_p^L\Big|_{series} 
= Z_{pp}^L - \frac{Z_{pn}^L Z_{np}^L}{Z_{nn}^S + Z_{nn}^L}
\nonumber \\
Z_p^S \Big|_{shunt} &= Z_p^S \Big|_{series} = Z_{pp}^S
\nonumber \\
Z_n^L\Big|_{shunt} &= Z_n^L\Big|_{series} 
= Z_{nn}^L - \frac{Z_{np}^L Z_{pn}^L}{Z_{pp}^S + Z_{pp}^L}
\nonumber \\
Z_n^S \Big|_{shunt} &= Z_n^S \Big|_{series} = Z_{nn}^S
\label{eq:Zp_oneMFD}
\end{align}

Three important observations are obtained from (\ref{eq:Zp_oneMFD}). As expected, in the source subsystem the original and modified impedances are equal, i.e. $Z_p^S = Z_{pp}^S$ because this subsystem is MFD. Second, in the load subsystem, the original sequence domain impedance no longer depends on injection type, i.e. $Z_p^L\Big|_{shunt} = Z_p^L\Big|_{series} $. The third observation is that $Z_p^L \neq Z_{pp}^L$. The difference between them is proportional to $Z_{pn}^L Z_{np}^L$, and also depends on the source impedance $Z_{nn}^S$. These observations can be seen in the simulation result shown Figure \ref{fig:Result_B_orig}.

\subsection{Special case in which both subsystems are MFD}
In this case, (\ref{eq:Zp_shunt}), (\ref{eq:Zp_series}) and (\ref{eq:additional_Z}) are reduced to:
\begin{align}
Z_p^L\Big|_{shunt} &= Z_p^L\Big|_{series} = Z_{pp}^L
\nonumber \\
Z_p^S\Big|_{shunt} &= Z_p^S\Big|_{series} = Z_{pp}^S
\nonumber \\
Z_n^L\Big|_{shunt} &= Z_n^L\Big|_{series} = Z_{nn}^L
\nonumber \\
Z_n^S\Big|_{shunt} &= Z_n^S\Big|_{series} = Z_{nn}^S
\label{eq:bothMFD}
\end{align} 

In other words, the original and modified sequence domain impedances are equal. This was also shown by (\ref{eq:Z_mfd}), and demonstrates the fact that MFD is a sufficient assumption for the original sequence domain impedances to be uniquely defined. The corresponding simulation result is shown in Figure \ref{fig:Result_C_orig}.

%
%

\end{document}